\documentclass[aps,nofootinbib,preprintnumbers]{revtex4}
\usepackage{lipsum}
\usepackage{CJK}
\usepackage{amsfonts}
\usepackage{amsmath}
\usepackage{amssymb}
\usepackage[english]{babel}
\usepackage{graphicx}
\usepackage{epsfig}
\usepackage{bm}
\usepackage{verbatim}
\usepackage[utf8]{inputenc}
\usepackage{booktabs}
\usepackage{multirow}
\usepackage{subfig}
\usepackage{slashed}
\usepackage{xcolor}
\usepackage[colorlinks=true,urlcolor=red,citecolor=red]{hyperref}
\usepackage[font=small]{caption}
\usepackage{float}
\usepackage{blindtext}
\usepackage{placeins}
\usepackage[utf8]{inputenc}
\usepackage{bbm}
\usepackage[colorinlistoftodos]{todonotes}
\newenvironment{Eqnarray}{\arraycolsep 0.14em\begin{eqnarray}}{\end{eqnarray}}
\newcommand{\ba}{\begin{Eqnarray}}
\newcommand{\ea}{\end{Eqnarray}}
\newcommand{\be}{\begin{equation}}
\newcommand{\ee}{\end{equation}}
\newcommand{\bal}{\begin{aligned}}
\newcommand{\eal}{\end{aligned}}
\newcommand{\bea}{\begin{eqnarray}}
\newcommand{\eea}{\end{eqnarray}}
\newcommand{\ben}{\begin{enumerate}}
\newcommand{\een}{\end{enumerate}}
\newcommand{\bit}{\begin{itemize}}
\newcommand{\eit}{\end{itemize}}
\newcommand{\bde}{\begin{widetext}}
\newcommand{\ede}{\end{widetext}}

\renewcommand{\[}{\left[}

\def\lsim{\mathrel{\rlap{\lower4pt\hbox{\hskip1pt$\sim$}}
    \raise1pt\hbox{$<$}}}
\def\gsim{\mathrel{\rlap{\lower4pt\hbox{\hskip1pt$\sim$}}
    \raise1pt\hbox{$>$}}}

\def\3211{$\mathrm{SU(3) \otimes SU(2)_L \otimes U(1)_R \otimes U(1)_{B-L}}$ }
\def\321{$\mathrm{SU(3) \otimes SU(2) \otimes U(1)}$ }
\def\422{$\mathrm{SU(4) \otimes SU(2) \otimes SU(2)_R}$ }

\newcommand{\mathsym}[1]{{}}

\newcommand{\ant}[1]{\textcolor{black}{#1}}
\newcommand{\ivo}[1]{\textcolor{black}{#1}}
\newcommand{\mar}[1]{\textcolor{black}{#1}}
\newcommand{\aur}[1]{\textcolor{black}{#1}} 
\pdfoutput=1
\begin{document}

\title{Controlled fermion mixing and FCNCs in a $\Delta(27)$ 3+1 Higgs Doublet Model 
%cobimaximal leptonic mixing pattern and radiative seesaw mechanism.
}

%Fermion masses and mixings and some phenomenological aspects in a
%with $\Delta \left( 27\right) $\ family symmetry.
\author{A. E. C\'{a}rcamo Hern\'{a}ndez$^{a}$}
\email{antonio.carcamo@usm.cl}
\author{Ivo de Medeiros Varzielas$^{b}$}
\email{ivo.de@udo.edu}
\author{M.L. L\'opez-Ib\'a\~{n}ez$^{c}$}
\email{maloi2@uv.es}
\author{Aurora Melis$^{d}$}
\email{aurora.melis@uv.es}
\date{\today }

\affiliation{
\vspace{2mm}
$^{{a}}$Universidad T\'{e}cnica Federico Santa Mar\'{\i}a\\ and Centro Cient\'{\i}fico-Tecnol\'{o}gico de Valpara\'{\i}so\\
Casilla 110-V, Valpara\'{\i}so, Chile,\\
$^{{b}}$ CFTP, Departamento de F\'{\i}sica, Instituto Superior T\'{e}cnico,\\
Universidade de Lisboa, Avenida Rovisco Pais 1, 1049 Lisboa, Portugal\\
$^{{c}}$ CAS Key Laboratory of Theoretical Physics, Institute of Theoretical Physics,\\ Chinese Academy of Sciences, Beijing 100190, China.\\
$^{{d}}$ 
Laboratory of High Energy and Computational Physics,\\ National Institute of Chemical Physics and Biophysics,\\ R\"avala pst. 10, 10143 Tallinn, Estonia\\
}

%-----------------------------------------------------
\begin{abstract}
We propose a 3+1 Higgs Doublet Model based on the $\Delta(27)$ family symmetry supplemented by several auxiliary cyclic symmetries leading to viable Yukawa textures for the Standard Model fermions, consistent with the observed pattern of fermion masses and mixings. The charged fermion mass hierarchy and the quark mixing pattern is generated by the spontaneous breaking of the discrete symmetries due to flavons that act as Froggatt-Nielsen fields. The tiny neutrino masses arise from a radiative seesaw mechanism at one loop level, thanks to a preserved $Z_2^{\left( 1\right)}$ discrete symmetry, which also leads to stable scalar and fermionic dark matter candidates. The leptonic sector features the predictive cobimaximal mixing pattern, consistent with the experimental data from neutrino oscillations. For the scenario of normal neutrino mass hierarchy, the model predicts an effective Majorana neutrino mass parameter in the range \mar{$3$~meV$\lesssim m_{\beta\beta}\lesssim$ $18$ meV}, which is within the declared range of sensitivity of modern experiments. The model predicts Flavour Changing Neutral Currents which constrain the model, for instance, $\mu\to e$ nuclear conversion processes and Kaon mixing are found to be within the reach of the forthcoming experiments.

\end{abstract}
%-----------------------------------------------------

%}
\maketitle
%-----------------------------------------------------
\section{Introduction \label{intro}}
%-----------------------------------------------------

The Standard Model (SM) is unable to describe the observed pattern of SM
fermion masses and mixings, which includes the large hierarchy among its numerous Yukawa couplings. To address the flavour problem, a promising option is to add
family symmetries and obtain the Yukawa couplings from an underlying theory
through the spontaneous breaking of the family symmetry.

$\Delta(27)$ as a family symmetry is greatly motivated by being one of the smallest discrete groups with a triplet and anti-triplet and the interesting interplay it has with CP symmetry.
$\Delta(27)$ has been used in \cite%
{Branco:1983tn,deMedeirosVarzielas:2006fc, Ma:2006ip,
Ma:2007wu,Bazzocchi:2009qg, deMedeirosVarzielas:2011zw, Varzielas:2012nn,Bhattacharyya:2012pi,Ferreira:2012ri,Ma:2013xqa,Nishi:2013jqa,Varzielas:2013sla,Aranda:2013gga,Varzielas:2013eta,Harrison:2014jqa,Ma:2014eka,Abbas:2014ewa,Abbas:2015zna,Varzielas:2015aua,Bjorkeroth:2015uou,Chen:2015jta,Vien:2016tmh,Hernandez:2016eod, Bjorkeroth:2016lzs, CarcamoHernandez:2017owh,deMedeirosVarzielas:2017sdv, Bernal:2017xat,CarcamoHernandez:2018iel,deMedeirosVarzielas:2018vab,CarcamoHernandez:2018hst,CarcamoHernandez:2018djj,Ma:2019iwj,Bjorkeroth:2019csz,CarcamoHernandez:2020udg}.

We consider here a 3+1 Higgs Doublet Model (HDM) based on the $\Delta(27)$ family symmetry supplemented by several cyclic symmetries, where three of the $SU(2)$ doublets transform as an
anti-triplet of $\Delta(27)$, $H$. The other doublet, $h$, does not acquire a
Vacuum Expectation Value (VEV) since it is charged under a preserved $Z_2^{\left( 1\right)}$ and couples only to the neutrino sector. Thus, the light active neutrino masses are generated from a radiative seesaw mechanism at one loop level mediated by the neutral components of the inert scalar doublet $h$ and the right handed Majorana neutrinos. Due to the preserved $Z_2^{\left( 1\right)}$ symmetry, our model has stable scalar and fermionic dark matter (DM) candidates. The scalar DM candidate is the lightest among the CP-even and CP-odd neutral components of the $SU(2)$-doublet scalar $h$. Furthemore, the fermionic DM candidate corresponds to the lightest among the right handed Majorana neutrinos. The DM constraints can be fulfilled in our model for an appropriate region of parameter space, along similar lines of Refs.~\cite{Diaz:2015pyv,Escudero:2016gzx,Arbelaez:2016mhg,Garcia-Cely:2015khw,Bernal:2017xat,Rojas-Abatte:2017hqm,Dutta:2017lny,Nomura:2017kih,CarcamoHernandez:2017kra,CarcamoHernandez:2017cwi,Gao:2018xld,Long:2018dun,CarcamoHernandez:2019cbd,Bhattacharya:2019fgs,Han:2019lux,CarcamoHernandez:2019xkb,CarcamoHernandez:2019lhv,CarcamoHernandez:2020ehn}. A detailed study of the implications of DM properties in our model goes beyond the scope of this paper and is therefore deferred for a future work. 
The masses and mixing of the
charged fermions arise from $H$. Realistic masses and mixing require further sources of $\Delta(27)$ breaking \cite{Bhattacharyya:2012pi,Varzielas:2013sla,Varzielas:2013eta} (this is not specific to $\Delta(27)$, see \cite{Felipe:2013vwa}). For this purpose, the model includes flavons
(singlets under the SM) that are triplets of $\Delta(27)$ and acquire VEVs
at a family symmetry breaking scale, assumed to be higher than the EW breaking
scale, thus allowing them to decouple from the low-energy scalar potential.

The Left-Handed (LH) leptons transform as anti-triplets of $\Delta(27)$, and the combination of
charged lepton couplings to $H$ and neutrino couplings to $h$ leads to a
model with radiative seesaw and featuring the predictive and viable cobimaximal mixing pattern, which has attracted a lot of attention and interest by the model building community due to its predictive power to yield the observed pattern of leptonic mixing \cite{Fukuura:1999ze,Miura:2000sx,Ma:2002ce,Ma:2015fpa,Ma:2016nkf,Damanik:2017jar,Ma:2017moj,Ma:2017trv,Grimus:2017itg,CarcamoHernandez:2017owh,CarcamoHernandez:2018hst,Ma:2019iwj,Ma:2019byo}.

The quarks transform as singlets of $\Delta(27)$ but their masses still
originate from Yukawa terms involving $H$ and a dominant flavon VEV. The symmetries
allow also terms with subdominant flavon VEVs which do not contribute to
the masses but do produce the leading contribution to Yukawa couplings
with the additional physical Higgs fields, and give rise to controlled Flavour Changing Neutral Currents (FCNCs).

Distinguishing family symmetry models that have similar predictions for the Yukawa couplings is particularly relevant, and FCNCs are arguably the most reliable way to do so (see e.g. \cite{Das:2016czs, Lopez-Ibanez:2017xxw, deMedeirosVarzielas:2018vab, Lopez-Ibanez:2019rgb} for some recent examples). 
In the present model, we study the FCNCs mediated by the physical scalars in the leptonic and quark sectors in order to constrain the parameter space, and find that in particular the muon conversion process and Kaon observables already constrain this model.

The layout of this paper is as follows. In Section \ref{sec:model} we describe the proposed model and we present its symmetry and field content.
Section \ref{sec:scalar} describes the low energy scalar potential and discusses the mass spectrum of the light scalars which play relevant roles in phenomenology.
In Section \ref{sec:fermion} we discuss the quark (\ref{sec:quark}) and lepton (\ref{sec:lepton}) couplings to the scalars, showing the respective Lagrangian terms, Yukawa matrices that arise after family symmetry breaking, and model's fits to the observables. Section \ref{sec:FCNC} analyses the constraints that arise from FCNCs in the context of this model. We conclude in Section \ref{sec:conclusion}.

%-----------------------------------------------------
\section{The model \label{sec:model}}
%-----------------------------------------------------
We consider an extension of the SM with additional family symmetry, which is
broken at a high scale. The full symmetry $\mathcal{G}$ of the model
exhibits the following spontaneous symmetry breaking pattern: 
\begin{eqnarray}
&&\mathcal{G}=SU(3)_{C}\times SU\left( 2\right) _{L}\times U\left( 1\right)
_{Y}\times \Delta \left( 27\right) \times Z_{2}^{\left( 1\right) }\times
Z_{2}^{\left( 2\right) }\times Z_{2}^{\left( 3\right) }\times Z_{18}  \notag
\\
&&\hspace{35mm}\Downarrow \Lambda  \notag \\[0.12in]
&&\hspace{15mm}SU(3)_{C}\times SU\left( 2\right) _{L}\times U\left( 1\right)
_{Y}\times Z_{2}^{\left( 1\right) }  \notag \\[0.12in]
&&\hspace{35mm}\Downarrow v  \notag \\[0.12in]
&&\hspace{15mm}SU(3)_{C}\times U\left( 1\right) _{Q}\times Z_{2}^{\left(
1\right) } \,,
\end{eqnarray}
where $\Lambda $ is the scale of breaking of the $\Delta \left( 27\right)
\times Z_{2}^{\left( 1\right) }\times Z_{2}^{\left( 2\right) }\times
Z_{2}^{\left( 3\right) }\times Z_{18}$ discrete group, which we assume to be
much larger than the electroweak symmetry breaking scale $v=246$ GeV. \ivo{The $Z_{18}$ symmetry and the three additional $Z_2$ symmetries are distinguished by superscripts and commute with $\Delta(27)$.}

The model includes four scalar $SU(2)_L$  doublets, three arranged as an anti-triplet of $\Delta(27)$, $H$, and $h$ which is a singlet of $\Delta(27)$, does not acquire a VEV, and is charged under the unbroken $%
Z_{2}^{\left( 1\right) }$. The scalar sector is further extended, to include four flavons (SM singlets) $%
\Delta(27)$ triplets $\phi_A$ and one $\Delta(27)$ trivial singlet $\sigma$
which plays the role of a Froggatt-Nielsen (FN) field. The FN field $\sigma$ acquires a VEV at a very large energy scale, spontaneously breaking the $Z_{18}$ discrete group and then giving rise to the observed SM fermion mass and mixing hierarchy. Furthermore, the $\Delta(27)$ triplet $\phi_3$ is introduced to build the quark Yukawa terms invariant under the $\Delta(27)$ family symmetry. The remaining $\Delta(27)$ triplets $\phi _{123}$, $\phi _{23}$ and $\phi _{1}$ are introduced in order to get a light active neutrino mass matrix featuring a cobimaximal mixing pattern, thus allowing to have a very predictive lepton sector consistent with the current neutrino oscillation experimental data. 
The scalar assignments under the $\Delta \left(27\right) \times Z_{2}^{\left( 1\right) }\times Z_{2}^{\left( 2\right)}\times Z_{2}^{\left( 3\right) }\times Z_{18}$ discrete group are shown in Table \ref{tab:scalars}. Here the dimensions of the $\Delta \left( 27\right) $ irreducible
representations are specified by the numbers in boldface and the different
charges are written in additive notation.

\begin{table}[t]
\begin{tabular}{|c|c|c|c|c|c|c|c|}\hline
    & $H$ & $h$ & $\sigma$ & $\phi _{123}$ & $\phi _{1}$ & $\phi _{23}$ & $\phi _{3}$ \\ \hline
    $\Delta\left( 27\right)$ & $\overline{\mathbf{3}}$ & $\mathbf{1}_{\mathbf{0,0}}$
    & $\mathbf{1}_{\mathbf{0,0}}$ & $\mathbf{3}$ & $\mathbf{3}$ & $\mathbf{3}$ & $\mathbf{3}$  \\ \hline
    $Z_{2}^{\left( 1\right) }$ & 0 & 1 & 0 & 0 & 0 & 0 & 0 \\ \hline
    $Z_{2}^{\left( 2\right) }$ & 0 & 0 & 0 & 0 & 0 & 1 & 0 \\ \hline
    $Z_{2}^{\left( 3\right) }$ & 0 & 0 & 0 & 0 & 0 & 0 & 1 \\ \hline
    $Z_{18}$ & 0 & 0 & -1 & 0 & 0 & 0 & 0\\ \hline
\end{tabular}
\caption{Scalar assignments under the $\Delta \left(27\right) \times Z_{2}^{\left( 1\right)} \times Z_{2}^{\left( 2\right)} \times Z_{2}^{\left( 3\right) }\times Z_{18}$. \ant{Superscripts differentiate between the multiple $Z_{2}$ symmetries.}}
\label{tab:scalars}
\end{table}
\begin{table}[h]
\begin{tabular}{|c|c|c|c|c|c|c|c|c|c|}\hline
& $q_{1L}$ & $q_{2L}$ & $q_{3L}$ & $u_{1R}$ & $u_{2R}$ & $u_{3R}$ & $d_{1R}$ & $d_{2R}$ & 
$d_{3R}$ \\ \hline
$\Delta\left( 27\right)$ & $\mathbf{1}_{\mathbf{0,0}}$ & $\mathbf{1}_{\mathbf{0,0%
}}$ & $\mathbf{1}_{\mathbf{0,0}}$ & $\mathbf{1}_{\mathbf{0,0}}$ & $\mathbf{1}_{%
\mathbf{0,0}}$ & $\mathbf{1}_{\mathbf{0,0}}$ & $\mathbf{1}_{\mathbf{0,0}}$ & 
$\mathbf{1}_{\mathbf{0,0}}$ & $\mathbf{1}_{\mathbf{0,0}}$ \\ \hline
$Z_{2}^{\left( 1\right) }$ & 0 & 0 & 0 & 0 & 0 & 0 & 0 & 0 & 0 \\ \hline
$Z_{2}^{\left( 2\right) }$ & 0 & 0 & 0 & 0 & 0 & 0 & 0 & 0 & 0 \\ \hline
$Z_{2}^{\left( 3\right) }$ & 0 & 0 & 0 & 1 & 1 & 1 & 1 & 1 & 1 \\ \hline
$Z_{18}$ & -4 & -2 & 0 & 4 & 2 & 0 & 4 & 3 & 3 \\ \hline
\end{tabular}
\caption{Quark assignments under the $\Delta \left(
27\right) \times Z_{2}^{\left( 1\right) }\times Z_{2}^{\left( 2\right)
}\times Z_{2}^{\left( 3\right) }\times Z_{18}$. \ant{Superscripts differentiate between the multiple $Z_{2}$ symmetries.}}
\label{tab:quarks}
\end{table}
\begin{table}[h]
\begin{tabular}{|c|c|c|c|c|c|c|c|}\hline
& $l_{L}$ & $l_{1R}$ & $l_{2R}$ & $l_{3R}$ & $N_{1R}$ & $N_{2R}$ & $N_{3R}$ \\ \hline
$\Delta \left( 27\right)$ & $\mathbf{\bar 3}$ & $\mathbf{1}_{\mathbf{0,1}}$
& $\mathbf{1}_{\mathbf{0,2}}$ & $\mathbf{1}_{\mathbf{0,1}}$ & $\mathbf{1}_{%
\mathbf{0,0}}$ & $\mathbf{1}_{\mathbf{0,0}}$ & $\mathbf{1}_{\mathbf{0,0}}$ \\ \hline
$Z_{2}^{\left( 1\right) }$ & 0 & 0 & 0 & 0 & 1 & 1 & 1 \\ \hline
$Z_{2}^{\left( 2\right) }$ & 0 & 0 & 0 & 0 & 0 & 0 & 1 \\ \hline
$Z_{2}^{\left( 3\right) }$ & 0 & 0 & 0 & 0 & 0 & 0 & 0 \\ \hline
$Z_{18}$ & 0 & 9 & 5 & 3 & 0 & 0 & 0\\ \hline
%\end{array}%
%$
\end{tabular}
\caption{Lepton assignments under the $\Delta \left(
27\right) \times Z_{2}^{\left( 1\right) }\times Z_{2}^{\left( 2\right)
}\times Z_{2}^{\left( 3\right) }\times Z_{18}$. \ant{Superscripts differentiate between the multiple $Z_{2}$ symmetries.}}
\label{tab:leptons}
\end{table}

The role of the different cyclic groups is described as follows. The $Z_{2}^{\left( 3\right) }$ symmetry is crucial for separating the $\Delta(27)$ scalar triplet $\phi_3$ participating in the quark Yukawa terms from the ones appearing in the neutrino Yukawa interactions. The $Z_{2}^{\left(2\right) }$ symmetry is necessary for shaping a cobimaximal texture of the light neutrino mass matrix, thus allowing a reduction of the lepton sector model parameters and at the same time allowing to successfully accommodate the neutrino oscillation experimental data. The preserved $Z_{2}^{\left(1\right) }$ symmetry allows the implementation of a radiative seesaw mechanism at one loop level, providing a natural explanation for the tiny masses of the light active neutrinos and also enabling stable DM candidates. Finally, the spontaneously broken $Z_{18}$ symmetry shapes a hierarchical structure of the SM charged fermion mass matrices which is crucial for a natural explanation of the SM charged fermion mass and quark mixing pattern.

The fermion sector includes three SM singlets, $Z_{2}^{\left( 1\right) }$
charged Right-Handed (RH) neutrinos $N_{iR}$ in addition to the SM fermions. All the fermions
are arranged as trivial singlets of $\Delta(27)$ with the exception of the
charged leptons fields, where the $SU(2)_L$ doublets $l_{L}$ transform as an
anti-triplet and the $l_{iR}$ transform as specific non-trivial singlets. The quark and lepton 
assignments under the $\Delta \left( 27\right) \times Z_{2}^{\left( 1\right)
}\times Z_{2}^{\left( 2\right) }\times Z_{2}^{\left( 3\right) }\times Z_{18}$
discrete group are shown in Tables \ref{tab:quarks} and \ref{tab:leptons}, respectively.

We stress here that, thanks to the preserved $Z_2^{\left( 1\right)}$ symmetry, the scalar and fermion sectors of our model contain stable DM candidates. The scalar DM candidate is the lightest among the CP-even and CP-odd neutral components of the $SU(2)$ scalar doublet $h$. The fermionic DM candidate corresponds to the lightest among the RH Majorana neutrinos. It is worth mentioning that in the scenario of a scalar DM candidate, it annihilates mainly into $WW$, $ZZ$, $t\overline{t}$, $b\overline{b}$ and $h_{\rm SM}h_{\rm SM}$ via a Higgs portal scalar interaction. These annihilation channels will contribute to the DM relic density, which can be accommodated for appropriate values of the scalar DM mass and of the coupling of the Higgs portal scalar interaction.  Thus, for the DM direct detection prospects, the scalar DM candidate would scatter off a nuclear target in a detector via Higgs boson exchange in the $t$-channel, giving rise to a constraint on the Higgs portal scalar interaction coupling. 
For the fermionic DM candidate, the lightest RH neutrino, the DM relic abundance can be obtained through freeze-in, as shown in \cite{Bernal:2017xat}.
 The DM constraints can therefore be fulfilled in our model for an appropriate region of parameter space, along similar lines of Refs.~\cite{Diaz:2015pyv,Escudero:2016gzx,Arbelaez:2016mhg,Garcia-Cely:2015khw,Bernal:2017xat,Rojas-Abatte:2017hqm,Dutta:2017lny,Nomura:2017kih,CarcamoHernandez:2017kra,CarcamoHernandez:2017cwi,Gao:2018xld,Long:2018dun,CarcamoHernandez:2019cbd,Bhattacharya:2019fgs,Han:2019lux,CarcamoHernandez:2019xkb,CarcamoHernandez:2019lhv,CarcamoHernandez:2020ehn,Cabrera:2020lmg}. A detailed study of the implications of the DM candidates in our model is nevertheless beyond the scope of this work.

With the particle content previously described, the scalar potential, as well as the Yukawa terms of up quarks, down quarks, charged leptons and the
neutrino terms are constrained by the symmetries, which we consider in detail in the following Sections.

%-----------------------------------------------------
\section{The low energy scalar potential \label{sec:scalar}}
%-----------------------------------------------------

The pattern of VEVs that we consider is

\begin{eqnarray}
\left\langle H\right\rangle &=&v_{H}\left( 0,0,1\right) ,\hspace{1.5cm}%
\left\langle \phi _{1}\right\rangle =v_{1}\left( 1,0,0\right) ,\hspace{1.5cm}%
\left\langle \phi _{3}\right\rangle =v_{3}\left( 0,0,1\right) , \\
\left\langle \phi _{123}\right\rangle &=&v_{123}\left( 1,\omega ,\omega
^{2}\right) ,\hspace{1cm}\left\langle \phi _{23}\right\rangle =v_{23}\left(
0,1,-1\right) ,  \hspace{1cm} \langle \sigma \rangle =v_{\sigma }\sim \lambda\,\Lambda\,,\label{eq:VEVpattern}
\end{eqnarray}\\
with $v_{H}=\frac{v}{\sqrt{2}}$, being $v=246$ GeV, and $\lambda\simeq 0.225$ the Cabibbo angle. We do not consider here in detail the potential terms that give rise to the flavon VEVs. The special $\Delta(27)$ VEV directions shown above and used in our model have been obtained in the literature in the framework of Supersymmetric models with $\Delta (27)$ family symmetry through D-term alignment mechanism \cite%
{deMedeirosVarzielas:2006fc} or F-term alignment mechanism \cite%
{Varzielas:2015aua}. Such VEV patterns have also been derived in non-supersymmetric models and have shown to be consistent with the scalar potential minimization equations for a large region of parameter space, as discussed in detail in \cite{Hernandez:2016eod,CarcamoHernandez:2017owh,CarcamoHernandez:2018hst} (see also \cite{Varzielas:2016zjc, deMedeirosVarzielas:2017glw}).

For the low energy scalar potential, we consider that the flavons have been integrated out, and write the scalar potential in four parts
\begin{equation}
V=V_{H}+V_{h}+V_{Hh}+V_{Hh}^{breaking} \,.
\end{equation}%

We write the $\Delta(27)$-invariant potential for $H$ in the notation of \cite{Varzielas:2016zjc, deMedeirosVarzielas:2017glw}
\begin{eqnarray}
V ( H ) &=& - ~\mu^2_{H}\sum_{i, \alpha} H_{i \alpha} H^{*i\alpha} + s
\sum_{i, \alpha, \beta} ( H_{i \alpha} H^{*i\alpha})( H_{i \beta}
H^{*i\beta})  \notag \\
&+& \sum_{i, j, \alpha, \beta} \left[ r_1 ( H_{i \alpha} H^{*i\alpha})( H_{j
\beta} H^{*j \beta}) + r_2 ( H_{i \alpha} H^{*i \beta})( H_{j \beta}
H^{*j\alpha}) \right]  \label{eq:potV0H}\\
&+& \sum_{\alpha, \beta} \left[d \left( H_{1 \alpha} H_{1 \beta} H^{*2
\alpha} H^{*3 \beta} + \text{cycl.} \right) + \text{h.c.}\right] \,, \notag
\end{eqnarray}
where the Greek letters denote the $SU(2)_L$ indices. An equivalent way of writing $V ( H )$ where the $\Delta(27)$ invariance is more transparent is shown in Appendix \ref{app:D27pot}.

The potential for the unbroken $Z_{2}^{\left( 1\right) }$-odd field $h$ is simply
\begin{equation} \label{eq:Vh}
V_{h}=\mu _{h}^{2}\left( hh^{\dagger }\right) +\gamma _{1}\left( hh^{\dagger
}\right) ^{2} \,,
\end{equation}
whereas the terms mixing $h$ and the $\Delta(27)$ triplet $H$ are
\begin{equation} \label{eq:VHh}
V_{Hh}=\alpha _{1}\left( HH^{\dagger }\right) _{\mathbf{\mathbf{1}_{0\mathbf{%
,0}}}}\left( hh^{\dagger }\right) +\alpha _{2}\left( \left( Hh^{\dagger
}\right) \left( H^{\dagger }h\right) \right) _{\mathbf{\mathbf{1}_{0\mathbf{%
,0}}}} \,,
\end{equation}
and expand to
\begin{equation}
V_{Hh}=\alpha _{1}\left( H_{1}H_{1}^{\dagger }+H_{2}H_{2}^{\dagger
}+H_{3}H_{3}^{\dagger }\right) \left( hh^{\dagger }\right) +\alpha _{2}\left[
\left( H_{1}h^{\dagger }\right) \left( H_{1}^{\dagger }h\right) +\left(
H_{2}h^{\dagger }\right) \left( H_{2}^{\dagger }h\right) +\left(
H_{3}h^{\dagger }\right) \left( H_{3}^{\dagger }h\right) \right] \,. 
\end{equation}

We also consider higher order terms allowed by the symmetries, even though they
are suppressed. In these, we find the leading order contribution to the mass splitting
between the CP-even and CP-odd neutral components of $h$, arises from
the terms:
\begin{equation} \label{eq:VNR}
V_{Hh}^{breaking}=\kappa_{1}\left[ \left( H^{\dagger }h\right) \left( H^{\dagger
}h\right) \right] _{\overline{\mathbf{3}}_{S_{1}}}\frac{\phi _{123}}{\Lambda 
}+\kappa_{2}\left[ \left( H^{\dagger }h\right) \left( H^{\dagger }h\right) %
\right] _{\overline{\mathbf{3}}_{S_{2}}}\frac{\phi _{123}}{\Lambda }+{\rm h.c.} \,.
\end{equation}
We present these terms as the splitting of the masses is needed in order to obtain viable neutrino masses through the radiative seesaw mechanism (see Section \ref{sec:lepton}).
Another invariant term arises by replacing $\phi _{123}$ by $\phi _{1}$, but
that term does not produce the effective mass term needed to yield the
mass splitting between the CP-even and CP-odd neutral components of $h$.
From the non-renormalizable scalar interactions given in Eq.~(\ref{eq:VNR}), using the corresponding $\Delta(27)$ breaking VEV, we obtain:
\begin{eqnarray}
V_{Hh}^{breaking} &=&\beta _{1}\left[ \left( H_{1}^{\dagger
}h\right) \left( H_{1}^{\dagger }h\right) + \omega \left( H_{2}^{\dagger }h\right)
\left( H_{2}^{\dagger }h\right) + \omega^2 \left( H_{3}^{\dagger }h\right) \left(
H_{3}^{\dagger }h\right) \right]   \notag \\
&+&\beta_{2}\left[ \left( H_{2}^{\dagger }h\right) \left(
H_{3}^{\dagger }h\right) + \omega \left( H_{1}^{\dagger }h\right) \left(
H_{3}^{\dagger }h\right) + \omega^2 \left( H_{1}^{\dagger }h\right) \left(
H_{2}^{\dagger }h\right) \right]+{\rm h.c.},\label{eq:VHhbreak}
\end{eqnarray}
where $\beta_{i}\equiv \kappa_i\, v_{123}/\Lambda$.
%------------------
\begin{figure}[t!]
    %\vspace{-1.cm}
    \centering
    \begin{minipage}{\textwidth}
        \includegraphics[width=0.4\textwidth]{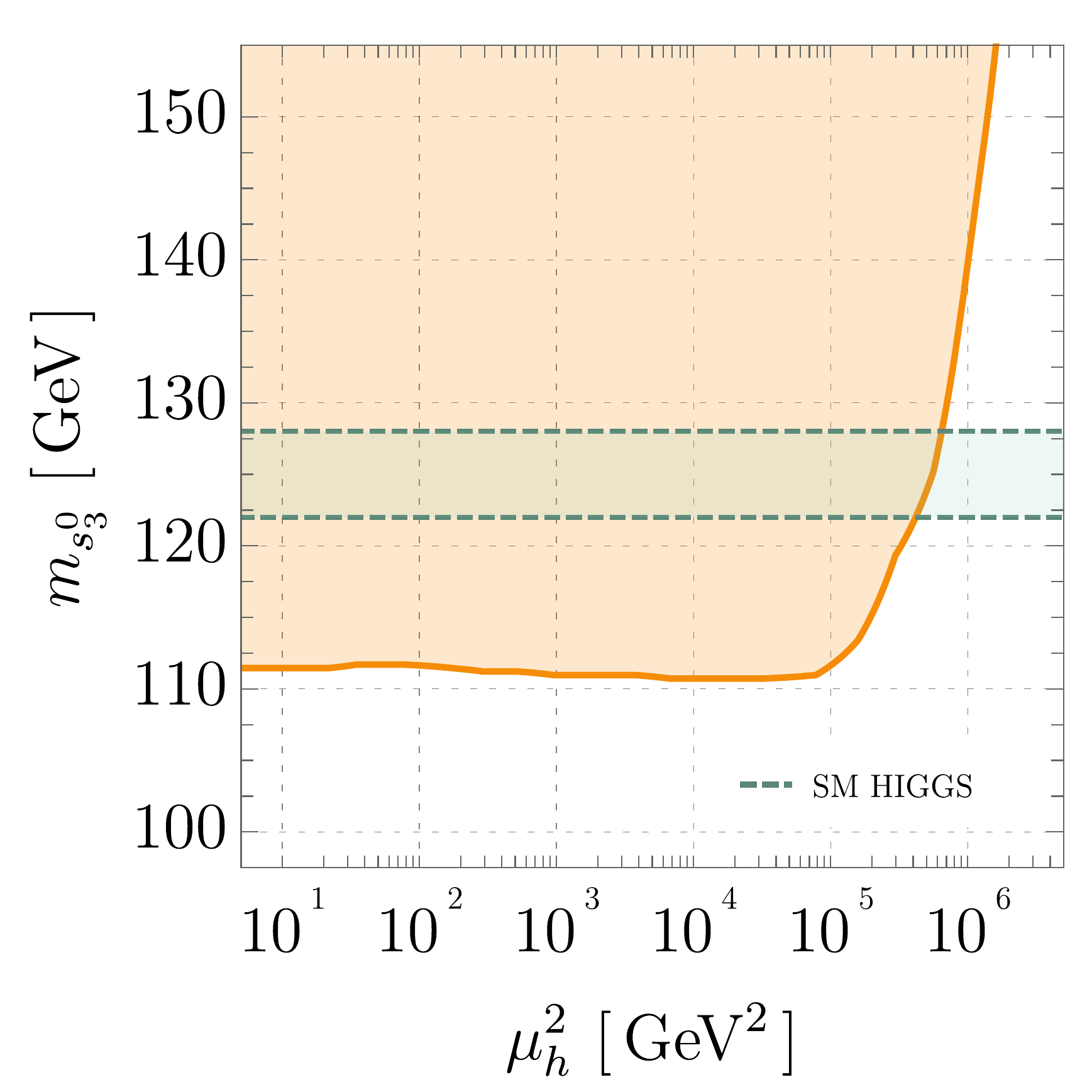}
        \hspace{0.5cm}
        \includegraphics[width=0.4\textwidth]{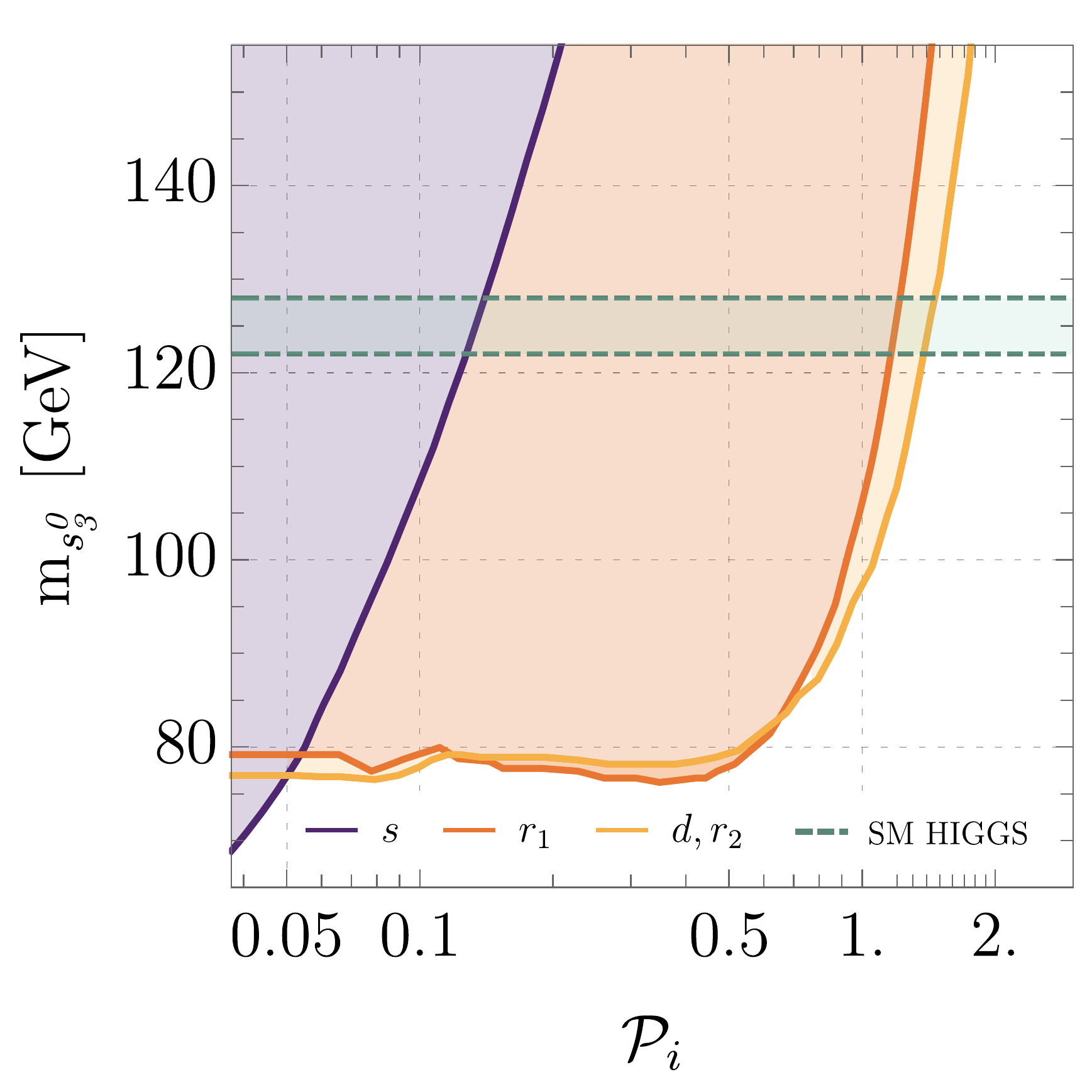}
    \end{minipage}\par\vspace{0.cm}
    \captionsetup{width=\textwidth}
    \caption{Loop corrected masses versus the parameters of the scalar potential \aur{in Eq~(\ref{eq:potV0H}): $\mu^2_h$ (Left) and $\mathcal{P}_i=s,r_1,r_2,d$ (Right)}.
    The rectangular green band corresponds to the allowed values of the $125$ GeV SM like Higgs boson; an uncertainty of $3$ GeV is assumed in the numerical computation of the Higgs mass.}
    \label{fig:mHth}
\end{figure}
%------------------

The electroweak symmetry is spontaneously broken by the non-zero VEV of the third component of the $\Delta(27)$ scalar triplet, $H_3$.
After that, three electrically charged and seven neutral Higgs fields arise.
The latter correspond to three CP-even ($s^0_1$, $s^0_2$, $s^0_3$), two CP-odd ($p^0_1$ and $p^0_2$) and two CP-mixed states ($h^0_1$ and $h^0_2$).
At tree-level, the light and heavy scalars and pseudoscalars, arising from the mixing of the neutral components of $H_1$ and $H_2$, are degenerate in mass:
\begin{equation} \label{eq:MSandMP}
    m_{s^0_1,\, s^0_2}^2 ~=~ m_{p^0_1,\, p^0_2}^2 ~=~ v^2\; \left(\, r_1 \,+\, r_2 \,-\, s \;\mp\; \frac{d}{2} \,\right),
\end{equation}
where the tadpole relation,
\begin{equation}
    \frac{dV}{dH^0_3} ~=~ -\mu^2_H\, v \;+\; s\, v^3 ~=~ 0 \,,
\end{equation}
has been taken into account.
As $H_3$ gets the non-zero VEV, it dos not mix with the first and second components of $H$.
As usual, its CP-odd and charged component are absorbed by the gauge bosons, which acquire masses, and a neutral massive scalar appears.
We identify it with the SM-like Higgs boson of mass $125$ GeV:
\begin{equation} \label{eq:MS3}
    m^2_{h_{\rm SM}} ~\equiv~ m_{s^0_3}^2 ~=~ 2\; s\; v^2.
\end{equation}
The CP-mixed neutral states are related to the $\Delta(27)$ singlet, whose squared mass matrix is:
\begin{equation}
    {\cal M}_h^2 ~=~ \begin{pmatrix}
                        ~\mu _{h}^{2} \,+\, \frac{v^2}{2}\, \left(\alpha_1 + \alpha_2 - \beta_1 \right) & \beta_1\, v^2\, \sin\frac{\pi}{3} \\[10pt] 
                        \beta_1\, v^2\, \sin \frac{\pi}{3} & \mu _{h}^{2} \,+\, \frac{v^2}{2} \left(\alpha_1 + \alpha_2  + \beta_1 \right)~ 
\label{Mh}                     \end{pmatrix} \,.
\end{equation}
As follows from Eq.~(\ref{Mh}), the mixing between the scalar and pseudoscalar components of $h$ is proportional to $\beta_1$ and, therefore, negligible if $\mu_h^2 \gg \beta_1\, v^2\, \sin \pi/3$.
At tree-level, the eigenmasses are
\begin{equation} \label{eq:Mh}
    m^2_{h^0_1,h^0_2} =\mu _{h}^{2}+\frac{v^2}{2}(\alpha _{1}+
\alpha_{2} \mp 2\,\beta_1).
    %\mu^2_h \;+\; \frac{v_H^2}{2}\, \bigg(\, \alpha_1 \,+\, \alpha_2 \;\mp\; 2\, \sqrt{\varepsilon_1^2 \,+\, \beta_1^2 \,-\, \varepsilon_1 \beta_1} \,\bigg).
\end{equation}

The phenomenology of the model is analysed by implementing it in SARAH 4.0.4 \cite{Staub:2008uz, Staub:2013tta, Porod:2014xia, Goodsell:2015ira, Staub:2015kfa, Goodsell:2018tti} and generating the corresponding SPheno code \cite{Porod:2003um, Porod:2011nf}, through which the numerical simulation in Section \ref{sec:FCNC} is performed.
In particular, loop corrections are taken into account to compute the spectrum of the model. They are specially important for the SM-like Higgs, $s^0_3$, whose mass is very sensitive to radiative corrections from other scalars.
In Figure \ref{fig:mHth}, the loop-corrected mass of this scalar is represented against the parameters of the scalar potential \aur{in Eq~(\ref{eq:potV0H}): $\mu^2_h, s, r_1, r_2 $ and $d$}.
The coloured regions correspond to the parameter space of our model.
The green band reflects a theoretical uncertainty of $3$ GeV that we consider in the estimation of the mass.
As it can be observed, the requirement of reproducing the $125$ GeV measured value sets non-trivial limits on some of the masses and quartic couplings in Eqs.\eqref{eq:potV0H} and \eqref{eq:Vh}:
\begin{equation}
    \mu^2_h\lesssim 6\times 10^5\, {\rm GeV}^2,\hspace{1.5cm}
    s\lesssim 0.14,\hspace{1.5cm}
    r_1\lesssim 1.2,\hspace{1.5cm}
    r_2,\, d \lesssim 1.5.
\end{equation}
The other parameters in Eqs.\eqref{eq:VHh} and \eqref{eq:VNR}, which are not bounded by the mass of the SM-like Higgs, are varied in the general range ${\cal P}_i\in [0.05,\, 2.]$ during the numerical scan.
Within those intervals, the masses of the resulting spectrum are:
\begin{equation}
    m_{s^0_1,\, p^0_1} \;\lsim\; 275\, {\rm GeV}, \hspace{1.cm}
    m_{s^0_2,\, p^0_2} \;\lsim\; 350\, {\rm GeV}, \hspace{1.cm}
    m_{h^0_1,\, h^0_2} \;\lsim\; 1\, {\rm TeV}.
\end{equation}

%-----------------------------------------------------
\section{Fermion masses and mixings \label{sec:fermion}}
%-----------------------------------------------------
%------------------
\subsection{Quark masses and mixings \label{sec:quark}}
%------------------
In the quark sector, due to the fields transforming
as $\Delta(27)$ trivial singlets, there are several terms as the nine possible
combinations of $\bar{q}_{iL} u_{jR}$ and the nine of $\bar{q}_{iL} d_{jR}$ are allowed by the symmetries.
The quarks must necessarily couple to $H$ because $h$ is secluded to the neutrino
sector through the unbroken $Z_{2}^{\left( 1\right) }$. We present first the quark terms that involve $H$ contracting with $\phi_3$, which eventually lead to the quark mass terms when the scalars acquire the respective VEVs ($H$ acquiring a VEV in the third direction only):
\begin{eqnarray}
\label{eq:LYQ}
\mathcal{L}^{(Q)}_Y &=&\big(\overline{q}_{1L}\,\,\overline{q}_{2L}\,\,\overline{q}_{3L}\big) 
\begin{pmatrix}
y_{11}^{\left(U\right)}\cfrac{\sigma^8}{\Lambda^8} & y_{12}^{\left(U\right)}\cfrac{\sigma^6}{\Lambda^6}& y_{13}^{\left(U\right)}\cfrac{\sigma^4}{\Lambda^4} \\
y_{21}^{\left(U\right)}\cfrac{\sigma^6}{\Lambda^6} & y_{22}^{\left(U\right)}\cfrac{\sigma^4}{\Lambda^4}& 
y_{23}^{\left(U\right)}\cfrac{\sigma^2}{\Lambda^2} \\
y_{31}^{\left(U\right)}\cfrac{\sigma^4}{\Lambda^4} & y_{23}^{\left(U\right)}\cfrac{\sigma^2}{\Lambda^2}& 
y_{33}^{\left(U\right)} \\
\end{pmatrix} \cfrac{1}{\Lambda}\left(\phi^*_3\widetilde{H}\right)_{\mathbf{1}_{\mathbf{0,0}}}
\begin{pmatrix}
u_{1R}\\u_{2R}\\u_{3R}
\end{pmatrix}\\[8pt]
&+&
\big(\overline{q}_{1L}\,\,\overline{q}_{2L}\,\,\overline{q}_{3L}\big) 
\begin{pmatrix}
y_{11}^{\left(D\right)}\cfrac{\sigma^7}{\Lambda^7} & y_{12}^{\left(D\right)}\cfrac{\sigma^6}{\Lambda^6}& y_{13}^{\left(D\right)}\cfrac{\sigma^6}{\Lambda^6} \\
y_{21}^{\left(D\right)}\cfrac{\sigma^6}{\Lambda^6} & y_{22}^{\left(D\right)}\cfrac{\sigma^5}{\Lambda^5}& 
y_{23}^{\left(D\right)}\cfrac{\sigma^5}{\Lambda^5} \\
y_{31}^{\left(D\right)}\cfrac{\sigma^4}{\Lambda^4} & y_{23}^{\left(D\right)}\cfrac{\sigma^3}{\Lambda^3}& 
y_{33}^{\left(D\right)}\cfrac{\sigma^3}{\Lambda^3} \\
\end{pmatrix} \cfrac{1}{\Lambda}\big(\phi_3H\big)_{\mathbf{1}_{\mathbf{0,0}}}
\begin{pmatrix}
d_{1R}\\d_{2R}\\d_{3R}
\end{pmatrix}+ {\rm h.c.} \quad .
\end{eqnarray}

The remaining quark terms have $H$ coupling to $\phi_{123}$ or $\phi_1$ (instead of coupling to $\phi_3$):
\begin{eqnarray}
\label{eq:deltaLYQ}
\delta \mathcal{L}_{Y}^{\left( Q\right) } &=&\big(\overline{q}_{1L}\,\,\overline{q}_{2L}\,\,\overline{q}_{3L}\big) 
\begin{pmatrix}
x_{11}^{\left(U\right)}\cfrac{\sigma^8}{\Lambda^8} & x_{12}^{\left(U\right)}\cfrac{\sigma^6}{\Lambda^6}& x_{13}^{\left(U\right)}\cfrac{\sigma^4}{\Lambda^4} \\
x_{21}^{\left(U\right)}\cfrac{\sigma^6}{\Lambda^6} & x_{22}^{\left(U\right)}\cfrac{\sigma^4}{\Lambda^4}& 
x_{23}^{\left(U\right)}\cfrac{\sigma^2}{\Lambda^2} \\
x_{31}^{\left(U\right)}\cfrac{\sigma^4}{\Lambda^4} & x_{23}^{\left(U\right)}\cfrac{\sigma^2}{\Lambda^2}& 
x_{33}^{\left(U\right)} \\
\end{pmatrix} \sum_{r=S_{1},S_{2},A} \cfrac{c^{\left(U\right)}_r}{\Lambda^2}\left(\phi_{123}\widetilde{H}\right)_{\overline{\mathbf{3}}_{r}}\phi_3
\begin{pmatrix}
u_{1R}\\u_{2R}\\u_{3R}
\end{pmatrix}\\[8pt]
&+&
\big(\overline{q}_{1L}\,\,\overline{q}_{2L}\,\,\overline{q}_{3L}\big) 
\begin{pmatrix}
x_{11}^{\left(D\right)}\cfrac{\sigma^7}{\Lambda^7} & x_{12}^{\left(D\right)}\cfrac{\sigma^6}{\Lambda^6}& x_{13}^{\left(D\right)}\cfrac{\sigma^6}{\Lambda^6} \\
x_{21}^{\left(D\right)}\cfrac{\sigma^6}{\Lambda^6} & x_{22}^{\left(D\right)}\cfrac{\sigma^5}{\Lambda^5}& 
x_{23}^{\left(D\right)}\cfrac{\sigma^5}{\Lambda^5} \\
x_{31}^{\left(D\right)}\cfrac{\sigma^4}{\Lambda^4} & x_{23}^{\left(D\right)}\cfrac{\sigma^3}{\Lambda^3}& 
x_{33}^{\left(D\right)}\cfrac{\sigma^3}{\Lambda^3} \\
\end{pmatrix} \sum_{r=S_{1},S_{2},A}\cfrac{c^{\left(D\right)}_{r}}{\Lambda^2}\big(\phi^*_{123} H\big)_{\mathbf{3}_{r}}\phi^*_3
\begin{pmatrix}
d_{1R}\\d_{2R}\\d_{3R}
\end{pmatrix}\\[6pt]
&+& \left( \phi _{123}\rightarrow \phi _{1}\right)
+ {\rm h.c.} \,, \notag
\end{eqnarray}
where the $r$ subscript denotes the possible $\Delta(27)$ representation and 
\begin{align}
\left( \phi _{123}\widetilde{H}\right) _{\overline{\mathbf{3}}%
_{S_{1}}} & \supset \left( \widetilde{H}_{1},\omega \widetilde{H}_{2},\omega
^{2}\widetilde{H}_{3}\right) v_{123} \,, 
&\left( \phi _{1}\widetilde{H}\right) _{\overline{\mathbf{3}}_{S_{1}}} \supset
\left( \widetilde{H}_{1},0,0\right) v_{1} \notag \,,\\
\left( \phi _{123}\widetilde{H}\right) _{\overline{\mathbf{3}}%
_{S_{2}}} &\supset \left( \omega \widetilde{H}_{3}+\omega ^{2}\widetilde{H}%
_{2},\omega ^{2}\widetilde{H}_{1}+\widetilde{H}_{3},\widetilde{H}_{2}+\omega 
\widetilde{H}_{1}\right) v_{123} \,, 
&\left( \phi _{1}\widetilde{H}\right) _{\overline{\mathbf{3}}_{S_{2}}} \supset
\left( 0,\widetilde{H}_{3},\widetilde{H}_{2}\right) v_{1} \, ,\\
\left( \phi _{123}\widetilde{H}\right) _{\overline{\mathbf{3}}_{A}} & \supset
\left( \omega \widetilde{H}_{3}-\omega ^{2}\widetilde{H}_{2},\omega ^{2}%
\widetilde{H}_{1}-\widetilde{H}_{3},\widetilde{H}_{2}-\omega \widetilde{H}%
_{1}\right) v_{123}\,,  
&\left( \phi _{1}\widetilde{H}\right) _{\overline{\mathbf{3}}_{A}} \supset
\left( 0,\widetilde{H}_{3},\widetilde{H}_{2}\right) v_{1}\,.\notag
\end{align}
Similar products arise from $(\phi_{123}^* H)_{{\mathbf{3}_r}}$ with the conjugation $\omega\leftrightarrow \omega^2$.

After symmetry breaking, these terms lead to another contribution to the masses (which can be absorbed into the previous terms, as the structure is exactly the same), but also to Yukawa couplings to the other components of $H$. In the absence of these terms, we would have in place a Natural Flavour Conservation mechanism as only $H_3$ couples to the quarks, and no FCNCs from the neutral scalars. But with these terms, we have Yukawa couplings to $H_1$ and $H_2$. While they have the same overall texture as the mass terms, they have different coefficients, and therefore are only approximately diagonalized when going to the mass basis of the quarks. They are therefore a source of FCNCs which is controlled by the symmetries.
Explicitly, the mass matrices and Yukawa couplings take the forms
\begin{equation}
\label{eq:MU_MD}
M_{U}=\frac{v}{\sqrt{2}} \frac{v_{3}}{\Lambda}\,\left( 
\begin{array}{ccc}
y_{11}^{\left( U\right) }\lambda ^{8} & y_{12}^{\left( U\right) }\lambda ^{6}
& y_{13}^{\left( U\right) }\lambda ^{4} \\ 
y_{21}^{\left( U\right) }\lambda ^{6} & y_{22}^{\left( U\right) }\lambda ^{4}
& y_{23}^{\left( U\right) }\lambda ^{2} \\ 
y_{31}^{\left( U\right) }\lambda ^{4} & y_{32}^{\left( U\right) }\lambda ^{2}
& y_{33}^{\left( U\right) }%
\end{array}%
\right) \hspace{0.5cm},\hspace{0.5cm}
M_{D}=\frac{v}{\sqrt{2}} \frac{v_{3}}{\Lambda}\,\left( 
\begin{array}{ccc}
y_{11}^{\left( D\right) }\lambda ^{7} & y_{12}^{\left( D\right) }\lambda ^{6}
& y_{13}^{\left( D\right) }\lambda ^{6} \\ 
y_{21}^{\left( D\right) }\lambda ^{6} & y_{22}^{\left( D\right) }\lambda ^{5}
& y_{23}^{\left( D\right) }\lambda ^{5} \\ 
y_{31}^{\left( D\right) }\lambda ^{4} & y_{32}^{\left( D\right) }\lambda ^{3}
& y_{33}^{\left( D\right) }\lambda ^{3}%
\end{array}%
\right) \quad ,
\end{equation}

\begin{equation}
\label{eq:YH2UD_1}
Y_{H_{1}^{0}}^{\left( U\right) }=\omega\,\lambda^{ (U)}_{H_1^0\rm}\left( 
\begin{array}{ccc}
x_{11}^{\left( U\right) }\lambda ^{8} & x_{12}^{\left( U\right) }\lambda ^{6}
& x_{13}^{\left( U\right) }\lambda ^{4} \\ 
x_{21}^{\left( U\right) }\lambda ^{6} & x_{22}^{\left( U\right) }\lambda ^{4}
& x_{23}^{\left( U\right) }\lambda ^{2} \\ 
x_{31}^{\left( U\right) }\lambda ^{4} & x_{32}^{\left( U\right) }\lambda ^{2}
& x_{33}^{\left( U\right) }%
\end{array}%
\right) \hspace{0.5cm},\hspace{0.5cm}
Y_{H_{1}^{0}}^{\left( D\right) }=\omega^2\,\lambda^{\rm (D)}_{H_1^0\rm}\left(
\begin{array}{ccc}
x_{11}^{\left( D\right) }\lambda ^{7} & x_{12}^{\left( D\right) }\lambda ^{6}
& x_{13}^{\left( D\right) }\lambda ^{6} \\ 
x_{21}^{\left( D\right) }\lambda ^{6} & x_{22}^{\left( D\right) }\lambda ^{5}
& x_{23}^{\left( D\right) }\lambda ^{5} \\ 
x_{31}^{\left( D\right) }\lambda ^{4} & x_{32}^{\left( D\right) }\lambda ^{3}
& x_{33}^{\left( D\right) }\lambda ^{3}%
\end{array}%
\right) \,, 
\end{equation}
\begin{equation}
\label{eq:YH2UD_2}
Y_{H_{2}^{0}}^{\left( U\right) }=\lambda^{\rm (U)}_{H_2^0\rm}\left( 
\begin{array}{ccc}
x_{11}^{\left( U\right) }\lambda ^{8} & x_{12}^{\left( U\right) }\lambda ^{6}
& x_{13}^{\left( U\right) }\lambda ^{4} \\ 
x_{21}^{\left( U\right) }\lambda ^{6} & x_{22}^{\left( U\right) }\lambda ^{4}
& x_{23}^{\left( U\right) }\lambda ^{2} \\ 
x_{31}^{\left( U\right) }\lambda ^{4} & x_{32}^{\left( U\right) }\lambda ^{2}
& x_{33}^{\left( U\right) }%
\end{array}%
\right) \hspace{0.5cm},\hspace{0.5cm}
Y_{H_{2}^{0}}^{\left( D\right) }=\lambda^{\rm (D)}_{H_2^0\rm}\left( 
\begin{array}{ccc}
x_{11}^{\left( D\right) }\lambda ^{7} & x_{12}^{\left( D\right) }\lambda ^{6}
& x_{13}^{\left( D\right) }\lambda ^{6} \\ 
x_{21}^{\left( D\right) }\lambda ^{6} & x_{22}^{\left( D\right) }\lambda ^{5}
& x_{23}^{\left( D\right) }\lambda ^{5} \\ 
x_{31}^{\left( D\right) }\lambda ^{4} & x_{32}^{\left( D\right) }\lambda ^{3}
& x_{33}^{\left( D\right) }\lambda ^{3}%
\end{array}%
\right) \,, 
\end{equation}
\begin{equation}
\label{eq:YH2UD_3}
\delta Y_{H_{3}^{0}}^{\left( U\right) }=\omega^2\lambda^{\rm (U)}_{H_3^0\rm}\left( 
\begin{array}{ccc}
x_{11}^{\left( U\right) }\lambda ^{8} & x_{12}^{\left( U\right) }\lambda ^{6}
& x_{13}^{\left( U\right) }\lambda ^{4} \\ 
x_{21}^{\left( U\right) }\lambda ^{6} & x_{22}^{\left( U\right) }\lambda ^{4}
& x_{23}^{\left( U\right) }\lambda ^{2} \\ 
x_{31}^{\left( U\right) }\lambda ^{4} & x_{32}^{\left( U\right) }\lambda ^{2}
& x_{33}^{\left( U\right) }%
\end{array}%
\right) \hspace{0.5cm},\hspace{0.5cm}
\delta Y_{H_{3}^{0}}^{\left( D\right) }=\omega\,\lambda^{\rm (D)}_{H_3^0\rm}\left( 
\begin{array}{ccc}
x_{11}^{\left( D\right) }\lambda ^{7} & x_{12}^{\left( D\right) }\lambda ^{6}
& x_{13}^{\left( D\right) }\lambda ^{6} \\ 
x_{21}^{\left( D\right) }\lambda ^{6} & x_{22}^{\left( D\right) }\lambda ^{5}
& x_{23}^{\left( D\right) }\lambda ^{5} \\ 
x_{31}^{\left( D\right) }\lambda ^{4} & x_{32}^{\left( D\right) }\lambda ^{3}
& x_{33}^{\left( D\right) }\lambda ^{3}%
\end{array}%
\right) \,, 
\end{equation}
where it is convenient to introduce the global effective couplings as
\begin{eqnarray}
\label{eq:YHUD_eff}
\lambda^{\rm (U,D)}_{H^0_1}= \frac{v_3\,v_{123}}{\sqrt{2}\Lambda^2}\left(c^{(U,D)}_{S_2}-c^{(U,D)}_{A}\right)  \quad,\quad 
\lambda^{\rm (U,D)}_{H^0_2}= \frac{v_3\,v_{123}}{\sqrt{2}\Lambda^2}\left[\left(c^{(U,D)}_{S_2}+c^{(U,D)}_{A}\right) + \frac{v_1}{v_{123}}\right] \quad,\quad \lambda^{\rm (U,D)}_{H^0_3}= \frac{v_3\,v_{123}}{\sqrt{2}\Lambda^2}\,c^{(U,D)}_{S_1}\,. \notag\\
\end{eqnarray}
We note again that the textures are the same for the Yukawa couplings and the mass matrices, but with different coefficients, such that the Yukawa couplings to $H_1$ and $H_2$ are not diagonalized in the quark mass basis.

\ivo{The quark masses and mixings are governed by the parameters $y^{(U,D)}_{ij}$. The parameters that govern the \ant{FCNCs} in the quark sector are $x^{(U,D)}_{ij}$, $c^{(U,D)}_{S_2,A}$ \ant{and they give subleading contributions to the SM quark mass matrices}. Given the structure of the Yukawa couplings we do not consider our model to be predictive in the quark sector, beyond accounting for the hierarchies between the masses.}
The physical observables of the quark sector, i.e., the quark masses, CKM
parameters and Jarskog invariant ~\cite{Xing:2019vks,Zyla:2020zbs} can be very well
reproduced in terms of natural parameters of order one.  This is shown in Table %
\ref{QuarkObs}, which for each observable, compares the model value with the respective experimental value.

\begin{table}[t]
\begin{center}
\begin{tabular}{c|l|l}
\hline\hline
Observable & Model value & Experimental value \\ \hline
$m_{u}[\mathrm{MeV}]$ & \quad $1.52$ & \quad $1.24\pm 0.22$ \\ \hline
$m_{c}[\mathrm{GeV}]$ & \quad $0.63$ & \quad $0.63\pm 0.02$ \\ \hline
$m_{t}[\mathrm{GeV}]$ & \quad $172.7$ & \quad $172.9\pm 0.4$ \\ \hline
$m_{d}[\mathrm{MeV}]$ & \quad $2.88$ & \quad $2.69\pm 0.19$ \\ \hline
$m_{s}[\mathrm{MeV}]$ & \quad $55.2$ & \quad $53.5\pm 4.6$ \\ \hline
$m_{b}[\mathrm{GeV}]$ & \quad $2.86$ & \quad $2.86\pm 0.03$ \\ \hline
$\sin \theta^{q}_{12}$ & \quad $0.22627$ & \quad $0.22650\pm 0.00048$ \\ \hline
$\sin \theta^{q}_{23}$ & \quad $0.04077$ & \quad $0.04053^{+0.00083}_{-0.00061}$ \\ \hline
$\sin \theta^{q}_{13}$ & \quad $0.00369$ & \quad $0.00361^{+0.00009}_{-0.00011}$ \\ \hline
$J_q$ & \quad $3.05\times 10^{-5}$ & \quad $\left(3.00^{+0.15}_{-0.09}\right)\times
10^{-5}$ \\ \hline
\end{tabular}%
\end{center}
\caption{Model and experimental values of the quark masses and CKM
parameters.}
\label{QuarkObs}
\end{table}

The model values above are obtained from the following benchmark point:
\begin{eqnarray}
M_U&=&
\left(
\begin{array}{ccc}
 -0.00111287 & 0.00224708 & 0.276781 \\
 0.00214193 & -0.621473 & -0.860806 \\
 0.0434745 & 0.849889 & 173.079 \\
\end{array}
\right){\rm GeV} \,, \notag \\
M_D&=&\left(
\begin{array}{ccc}
 0.00396153 & 0.0120505 & -0.000101736-0.0100846 i \\
 0.00331057 & 0.0648106 & 0.0952388 \\
 -0.0480789 & 0.325728 & 2.82949 \\
\end{array}
\right){\rm GeV} \,.
\end{eqnarray}

%------------------
\subsection{Lepton masses and mixings \label{sec:lepton}}
%------------------
In the lepton sector, the number of Yukawa terms is much smaller due to the assignments under $\Delta(27)$. The charged lepton and neutrino Yukawa terms invariant under the symmetries of the model are given by:
\begin{eqnarray}
\label{eq:LYl}
\mathcal{L}_{Y}^{\left( l\right) } &=& y_{1}^{\left( l\right) }\frac{\sigma ^{9}}{\Lambda
^{9}} \left( \overline{%
l}_{L}H\right) _{\mathbf{1}_{\mathbf{0,2}}}l_{1R}+y_{2}^{\left( l\right) } \frac{\sigma ^{5}}{\Lambda ^{5}} \left( \overline{l}_{L}H\right) _{\mathbf{1}_{%
\mathbf{0,1}}}l_{2R}+y_{3}^{\left( l\right)
}\frac{%
\sigma ^{3}}{\Lambda ^{3}}\left( \overline{l}_{L}H\right) _{\mathbf{1}_{\mathbf{0,0}}}l_{3R}+{\rm h.c.}  \label{eq:Lyl}\quad ,\\[18pt]
\label{eq:LYnu}
\mathcal{L}_{Y}^{\left( \nu \right) } &=& y_{1}^{\left( \nu \right) }\frac{1}{\Lambda }\left( 
\overline{l}_{L}\phi _{1}^{*}\widetilde{h}\right) _{\mathbf{1}_{\mathbf{0,0}%
}}N_{1R}+y_{2}^{\left( \nu \right) }\frac{1}{\Lambda }\left( \overline{l}%
_{L}\phi_{123}^{*} \widetilde{h}\right) _{\mathbf{1}_{\mathbf{0,0}}}N_{2R} \label{Lynu} \\
&+& y_{3}^{\left( \nu \right) }\frac{1}{\Lambda }\left( \overline{l}_{L}\phi _{123}^{*}\widetilde{h}%
\right) _{\mathbf{1}_{\mathbf{0,0}}}N_{1R}+y_{4}^{\left(
\nu \right) }\frac{1}{\Lambda }\left( \overline{l}_{L}\phi _{1}^{*} \widetilde{h}\right) _{\mathbf{1%
}_{\mathbf{0,0}}}N_{2R}+y_{5}^{\left( \nu \right) }\frac{1}{\Lambda }\left( 
\overline{l}_{L}\phi _{23}^{*}\widetilde{h}\right) _{\mathbf{1}_{\mathbf{0,0}%
}}N_{3R}\\[5pt]
\label{eq:LMnu}
&+& \frac{m_{N_{1}}}{2}\overline{N}_{1R}N_{1R}^{c}+\frac{m_{N_{2}}}{2}\overline{N}%
_{2R}N_{2R}^{c}+\frac{m_{N_{3}}}{2}\overline{N}_{3R}N_{3R}^{c}+\frac{m_{N_{4}}}{2}\left(
\overline{N}_{1R}N_{2R}^{c}+\overline{N}_{2R}N_{1R}^{c}\right) +{\rm h.c.}  \quad ,
\end{eqnarray}%
where the dimensionless couplings in Eqs.~(\ref{eq:Lyl})-(\ref{Lynu}) are $%
\mathcal{O}(1)$ parameters.

From the charged lepton terms and the VEV pattern we consider (see Eq.~(\ref{eq:VEVpattern})), we obtain a diagonal mass matrix:
\begin{equation}
M_{l}=\left( 
\begin{array}{ccc}
m_{e} & 0 & 0 \\ 
0 & m_{\mu} & 0 \\ 
0 & 0 & m_{\tau}%
\end{array}%
\right).  \label{Mediag}
\end{equation}
with the charged lepton masses given by: 
\begin{equation}
m_{e}=y_{1}^{\left( l\right) }\frac{v\,v_{\sigma }^{9}}{\sqrt{%
2}\,\Lambda ^{9}}=y_{1}^{\left( l\right) }\lambda ^{9}\frac{v}{\sqrt{2}},%
\hspace{1cm}m_{\mu }=y_{2}^{\left( l\right) }\frac{v\,v_{\sigma
}^{5}}{\sqrt{2}\,\Lambda^{5}}=y_{2}^{\left( l\right) }\lambda ^{5}%
\frac{v}{\sqrt{2}},\hspace{1cm}m_{\tau}=y_{3}^{\left( l\right) }\frac{%
v\,v_{\sigma }^{3}}{\sqrt{2}\,\Lambda^{3}}=y_{3}^{\left(
l\right) }\lambda ^{3}\frac{v}{\sqrt{2}}\,,
\end{equation}%
where in a slight abuse of notation, we have absorbed the $\mathcal{O}(1)$ parameters of the VEVs into redefinitions of $y_{1}^{\left( l\right) }$, $y_{2}^{\left( l\right) }$ and $y_{3}^{\left( l\right) }$ in the expressions with $\lambda$. $y_{1}^{\left( l\right) }$, $y_{2}^{\left( l\right) }$ and $y_{3}^{\left( l\right) }$ are assumed to be real. 
As in the quark sector, the Lagrangian in Eq.\eqref{eq:LYl} gives rise to FCNCs through additional Yukawa couplings that arise with the other components of $H$, namely $H_1$ and $H_2$.
The entries are of the same size of those in $Y^{(l)}_{H_3}$ but in different positions, thus
\begin{equation}
    Y^{(l)}_{H^0_1} ~=~ \frac{\sqrt{2}}{v} \begin{pmatrix}
                        ~0~ & ~0~ & ~m_\tau~ \\
                        ~m_e~ & ~0~ & ~0~ \\
                        ~0~ & ~m_\mu~ & ~0~
                      \end{pmatrix},
    \hspace{1.cm}
    Y^{(l)}_{H^0_2} ~=~ \frac{\sqrt{2}}{v} \begin{pmatrix}
                        ~0~ & ~m_\mu~ & ~0~ \\
                        ~0~ & ~0~ & ~m_\tau~ \\
                        ~m_e~ & ~0~ & ~0~
                      \end{pmatrix}  .  
\end{equation}

\begin{figure}[t]
\includegraphics[width=0.47\textwidth]{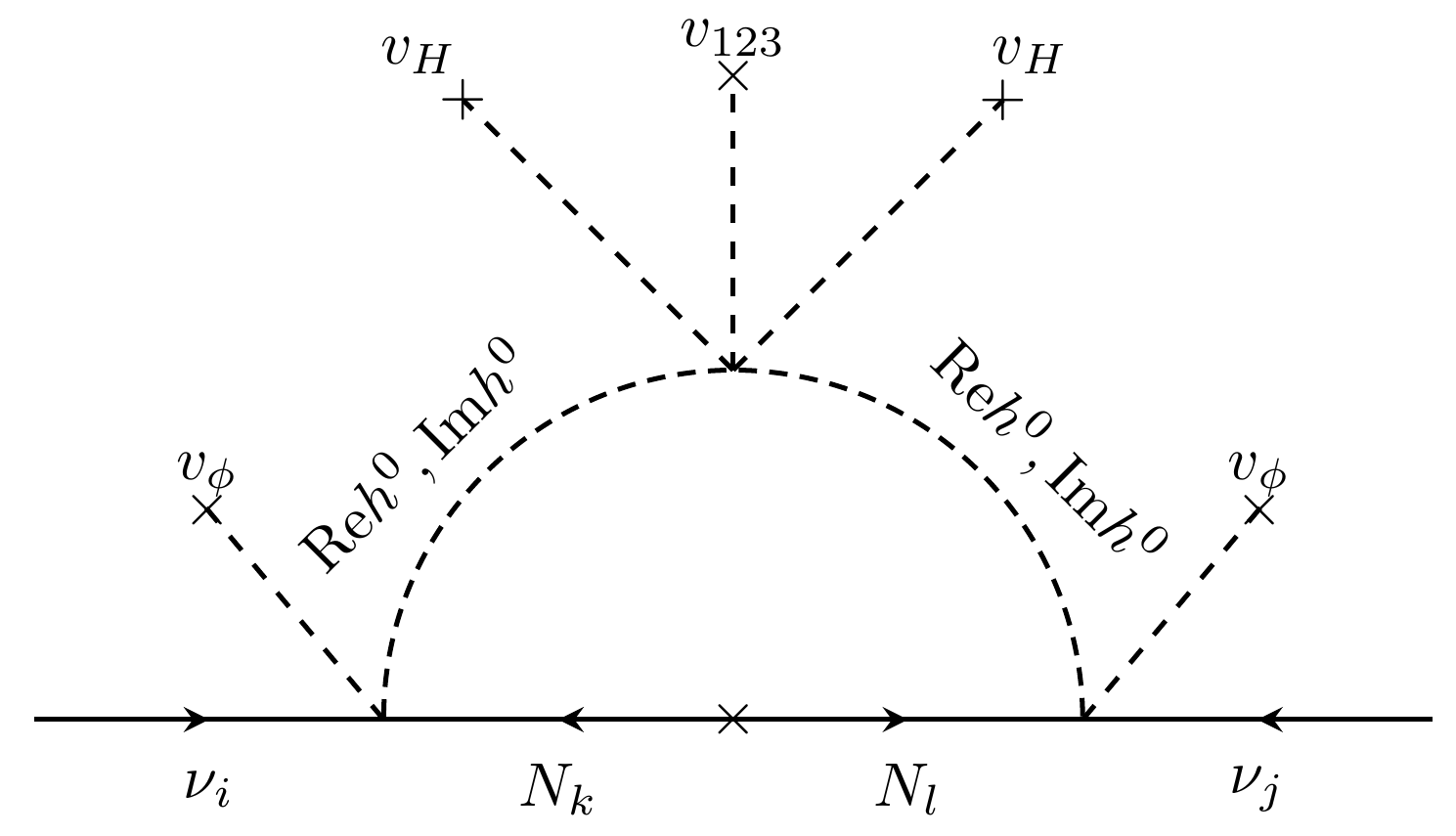} 
\caption{One-loop Feynman diagram contributing to the light active neutrino mass matrix. Here $i,j,k,l=1,2,3$ and $v_{\phi}$ stands for either $v_1$, $v_{23}$ or $v_{123}$.}
\label{Loopdiagramsmu}
\end{figure}

In the neutrino sector, sorting out the products in Eq.~(\ref{eq:LYnu}-\ref{eq:LMnu}), the Yukawa and Majorana mass matrices display the following structures:
\begin{eqnarray} \label{eq:Lv}
     Y_{\nu} = 
     \frac{1}{\sqrt{2}\Lambda}\,\begin{pmatrix}
     y^{(\nu)}_3 v_{123} + y^{(\nu)}_1 v_{1}  &  y^{(\nu)}_2 v_{123} + y^{(\nu)}_4 v_{1}& 0\\
     \omega^2\, y^{(\nu)}_3\,v_{123}& \omega^2\,y^{(\nu)}_2\,v_{123}& ~~y^{(\nu)}_5\,v_{23}\\
     \omega\,y^{(\nu)}_3 \,v_{123} & \omega\,y^{(\nu)}_2\,v_{123} & -y^{(\nu)}_5\,v_{23} \\
     \end{pmatrix}
     \qquad , \qquad
     M_N = 
     \begin{pmatrix}
     m_{N_1} & m_{N_4}& 0\\
     m_{N_4} &  m_{N_2} & 0\\
     0 & 0 & m_{N_3} \\
     \end{pmatrix}\,.
\end{eqnarray}

After the spontaneous breaking of the discrete symmetries and of the electroweak symmetry, the following neutrino Yukawa interactions arise:
\begin{eqnarray}
\mathcal{L}_{Y}^{\left( \nu \right) } &=&z_{1}^{\left( \nu \right) }%
\overline{\nu }_{1L}\left( h_{R}^{0}-i\,h_{I}^{0}\right) \widetilde{N}_{1R}+z_{2}^{\left(
\nu \right) }\left( \omega ^{2}\,\overline{\nu }_{2L}+\omega\,\overline{\nu }%
_{3L}\right) \left( h_{R}^{0}-i\,h_{I}^{0}\right) \widetilde{N}_{1R}\notag \\
&+&z_{3}^{\left( \nu \right) }\overline{\nu }_{1L}\left(
h_{R}^{0}-i\,h_{I}^{0}\right) \widetilde{N}_{2R}+z_{4}^{\left( \nu \right) }\left( \omega
^{2}\,\overline{\nu }_{2L}+\omega\,\overline{\nu }_{3L}\right) \left(
h_{R}^{0}-i\,h_{I}^{0}\right) \widetilde{N}_{2R}+z_{5}^{\left( \nu \right) }\left( \overline{\nu }_{2L}-\overline{\nu }%
_{3L}\right) \left( h_{R}^{0}-i\,h_{I}^{0}\right) N_{3R}\notag \\
&+&\frac{m_{\widetilde{N}_{1}}}{2}\overline{\widetilde{N}}_{1R}\widetilde{N}_{1R}^{c}+\frac{m_{\widetilde{N}_{2}}}{2}\overline{\widetilde{N}}%
_{2R}\widetilde{N}_{2R}^{c}+\frac{m_{N_{3}}}{2}\overline{N}_{3R}N_{3R}^{c}+\mathrm{h.c.}\,,
\end{eqnarray}
with $m_{h_{R}^{0}}=m_{{\rm Re}[h^{0}]}$ and $m_{h_{I}^{0}}=m_{{\rm Im}[h^{0}]}$, while $\widetilde{N}_{1R}$, $\widetilde{N}_{2R}$ are the physical Majorana neutrino fields arising from the combinations of $N_{1R}$ and $N_{2R}$. They are given by:
\begin{equation}
\label{eq:N12tilde}
\left( 
\begin{array}{c}
\widetilde{N}_{1R} \\ 
\widetilde{N}_{2R}%
\end{array}%
\right) =\left( 
\begin{array}{cc}
\cos \beta  & -\sin \beta  \\ 
\sin \beta  & \cos \beta 
\end{array}%
\right) \left( 
\begin{array}{c}
N_{1R} \\ 
N_{2R}%
\end{array}%
\right) 
\end{equation}
where the mixing angle $\beta$ takes the form $\tan 2\beta =-2m_{N_{4}}/(m_{N_{1}}-m_{N_{2}})$. The $z^{(\nu)}_i$ are the Yukawa parameters in the basis of diagonal $M_N$ obtained by performing the rotation in Eq.(\ref{eq:N12tilde}). In that basis, the neutrino Yukawa matrix ($Y_\nu \rightarrow \widetilde{Y}_\nu$) maintains the structure of Eq.\eqref{eq:Lv} but with new entries determined by $z^{(\nu)}_i$. The explicit expression for $\widetilde{Y}_\nu$ and the relation between the $y^{(\nu)}_i$ and $z^{(\nu)}_i$ parameters is given in Appendix~\ref{app:Mnupar}. %The relation with the $y^{\nu}_i$ is explicitly shown in Appendix~\ref{app:Mnupar}.
Therefore, in the basis where the RH neutrinos are diagonal%($Y_\nu \rightarrow \widetilde{Y}_\nu$)
, the light active neutrino mass matrix is obtained from the radiative seesaw mechanism as shown in the Feynman diagram of Figure \ref{Loopdiagramsmu} and it is given by:
\begin{eqnarray}
\label{eq:RADseesaw}
    M_{\nu} \equiv \frac{1}{2(4\pi)^2}\,\widetilde{Y}_{\nu} \,
    \begin{pmatrix}
    m_{\widetilde{N}_1}\,f_1 & 0 & 0\\
    0 & m_{\widetilde{N}_2}\,f_2 & 0\\
    0 & 0 & m_{N_3}\,f_3 
    \end{pmatrix}
    \,  \widetilde{Y}^T_{\nu} \,,
\end{eqnarray} 
with 
\begin{equation}
\label{eq:fk}
f_{k}=f\left( m_{h_{R}^{0}},m_{h_{I}^{0}},m_{\widetilde{N}_{k}}\right) ,\hspace{1.5cm}%
f_{3}=f\left( m_{h_{R}^{0}},m_{h_{I}^{0}},m_{N_{3}}\right) ,\hspace{1.5cm}%
k=1,2\quad.
\end{equation}
The loop function $f$ takes the form:
\begin{equation}
f\left( m_{h_{R}^{0}},m_{h_{I}^{0}},m_{N_{R}}\right) = \frac{m_{h_{R}^{0}}^{2}%
}{m_{h_{R}^{0}}^{2}-m_{N_{R}}^{2}}\ln \left( \frac{m_{h_{R}^{0}}^{2}}{%
m_{N_{R}}^{2}}\right) -\frac{m_{h_{I}^{0}}^{2}}{%
m_{h_{I}^{0}}^{2}-m_{N_{R}}^{2}}\ln \left( \frac{m_{h_{I}^{0}}^{2}}{%
m_{N_{R}}^{2}}\right) \,,
\end{equation}
where $m^2_{h^0_R},m^2_{h^0_I}$ are given in terms of the parameters of the scalar potential in the entries (1,1) and (2,2) of Eq.~(\ref{eq:Mh}).
One can show that the resulting light active
neutrino mass matrix in Eq.~(\ref{eq:RADseesaw}) can be parametrized as: 
\begin{equation}
\label{eq:Mnu}
M_{\nu}=\left( 
\begin{array}{ccc}
a & d \omega^{2} & d \omega  \\ 
d \omega^{2} & b e^{i\theta } & c \\ 
d \omega & c & b e^{-i\theta }%
\end{array}%
\right) \,,
\end{equation}
where the exact relations between the effective parameters $a,b,c,d,\theta$ and the lagrangian parameters $z^{(\nu)}_i$ are given in Appendix \ref{app:Mnupar}. Here, we stress that $c$ can be expressed in terms of $b$ and $\theta$, and that all the effective parameters depend on the flavon VEVs.

\begin{figure}[p]
\centering
\includegraphics[width=0.75\textwidth]{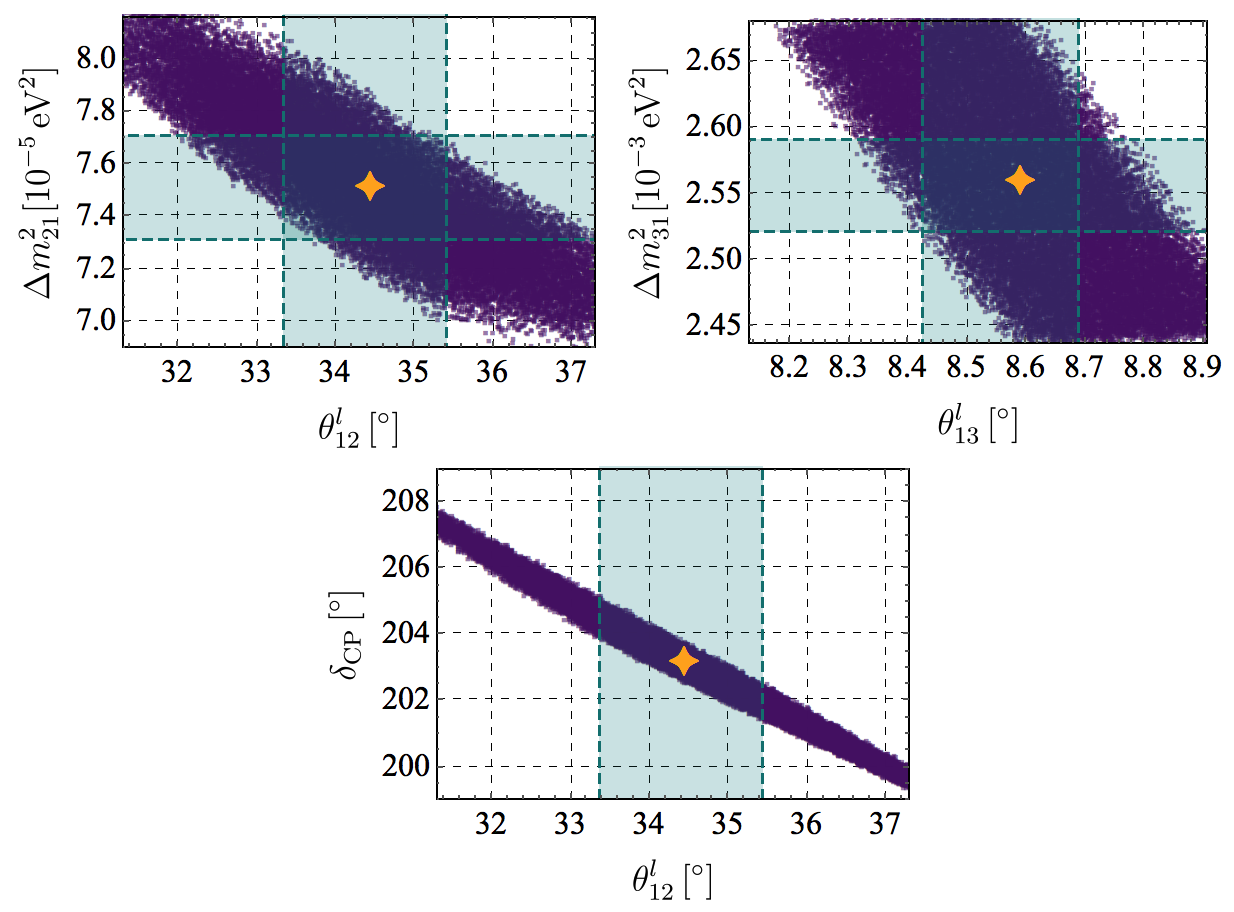}
\captionsetup{width=0.62\textwidth%, labelsep=none
    }
\caption{Correlation between neutrino observables around the benchmark point. The star corresponds to the benchmark point considered in the text, whereas the dashed lines correspond to the experimental $1 \sigma$ ranges of \cite{deSalas:2020pgw}.}
\label{fig:correlationleptonsector}
\vspace{3mm}
\includegraphics[width=0.65\textwidth]{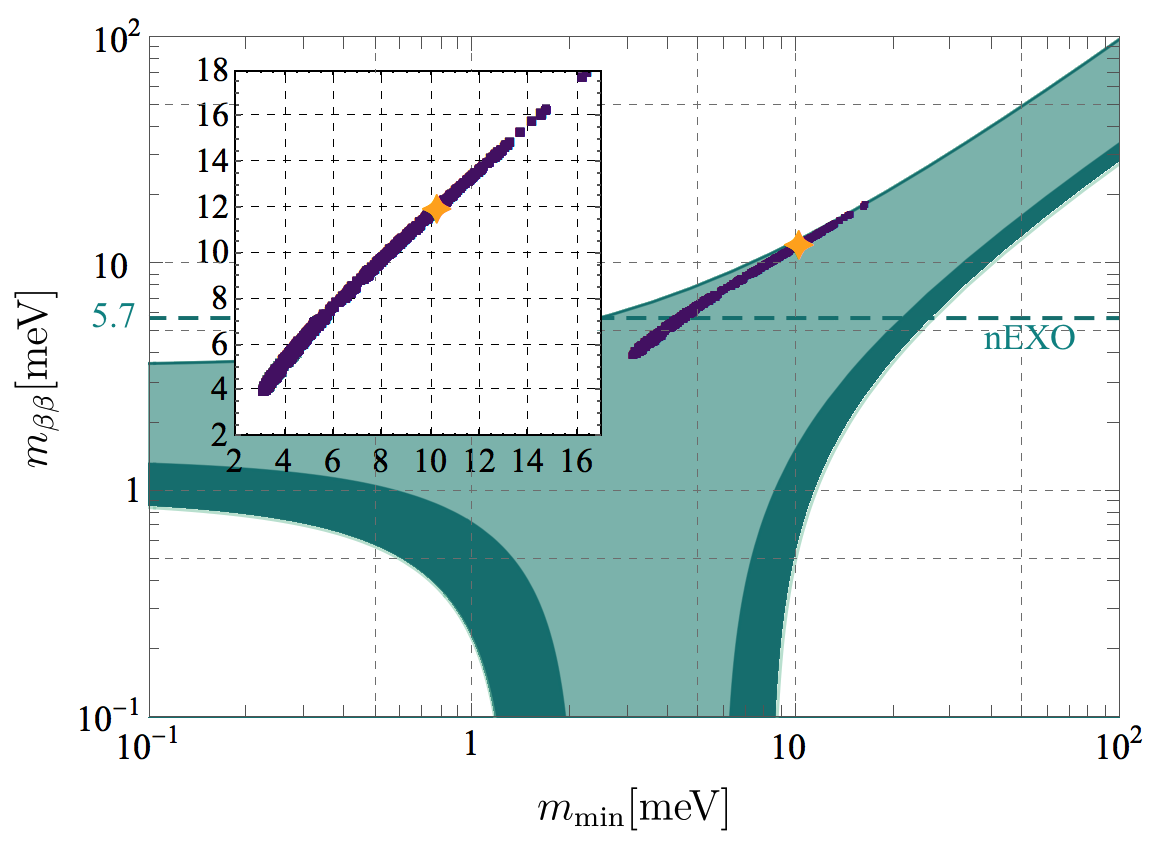}
\captionsetup{width=0.63\textwidth%, labelsep=none
}
\label{fig:mbb_vs_mminn}
    \caption{The effective Majorana neutrino mass parameter $m_{\beta\beta}$ against $m_{\rm min}=m_{\nu 1}$. The green shadow is the allowed for NO scenarios. The dashed line corresponds to the future sensitivity expected from the nEXO experiment. The inner plot zooms in the correlation between $m_{\beta\beta}$ and $m_{\rm min}$. }
\end{figure}
\begin{table}[t]
{\footnotesize \ }
\par
\begin{center}
{\footnotesize \ \renewcommand{\arraystretch}{1} 
\begin{tabular}{c|c||c|c|c|c}
\hline
\multirow{2}{*}{Observable} & \multirow{2}{*}{\parbox{7em}{Model\\ value} }
& \multicolumn{4}{|c}{Neutrino oscillation global fit values (NH)} \\ 
\cline{3-6}
&  & Best fit $\pm 1\sigma$ \cite{deSalas:2020pgw} & Best fit $\pm 1\sigma$ 
\cite{Esteban:2020cvm} & $3\sigma$ range \cite{deSalas:2020pgw} & $3\sigma$
range \cite{Esteban:2020cvm} \\ \hline\hline
$\Delta m_{21}^{2}$ [$10^{-5}$eV$^{2}$] & $7.51$ & $7.50_{-0.20}^{+0.22}$ & $%
7.42_{-0.20}^{+0.21}$ & $6.94-8.14$ & $6.82-8.04$ \\ \hline
$\Delta m_{31}^{2}$ [$10^{-3}$eV$^{2}$] & $2.56$ & $2.56_{-0.04}^{+0.03}$ & $%
2.517_{-0.028}^{+0.026}$ & $2.46-2.65$ & $2.435-2.598$ \\ \hline
$\theta^{l}_{12}[^{\circ }]$ & $34.45$ & $34.3\pm 1.0$ & $%
33.44_{-0.74}^{+0.77}$ & $31.4-37.4$ & $31.27-35.86$ \\ \hline
$\theta^{l}_{13}[^{\circ }]$ & $8.59$ & $8.58_{-0.15}^{+0.11}$ & $8.57\pm 0.12$
& $8.16-8.94$ & $8.20-8.93$ \\ \hline
$\theta^{l}_{23}[^{\circ }]$ & $44.89$ & $48.79_{-1.25}^{+0.93}$ & $%
49.2_{-1.2}^{+0.9} $ & $41.63-51.32$ & $40.1-51.7$ \\ \hline
$\delta_{\rm CP}[^{\circ }]$ & $203.15$ & $216_{-25}^{+41}$ & $197_{-24}^{+27}$ & $%
144-360$ & $120-369$ \\ \hline\hline
\end{tabular}
{\normalsize \ }}
\end{center}
\par
{\footnotesize \ }
\caption{Model and experimental values of the neutrino mass squared
splittings, leptonic mixing angles, and CP-violating phase. The
experimental values are taken from Refs.~\protect\cite%
{deSalas:2020pgw,Esteban:2020cvm}.}
\label{tab:neutrinos-NH}
\end{table}

\ivo{The masses of the charged leptons are set to the observed values, and the remaining parameters that govern the neutrino sector are $y^{(\nu)}_{1,2,3,4,5}$, $m_{N_{1,2,3,4}}$. The model is predictive as only the combinations $a, b, c, d, \theta$ of these parameters affect the}
physical observables of the neutrino sector, i.e., the three leptonic
mixing angles, the CP phase and the neutrino mass squared splittings for the
normal mass hierarchy (NH). \ivo{These observables} can be very well reproduced, as shown in Table %
\ref{tab:neutrinos-NH}, starting from the following benchmark point: 
\begin{equation}
a\simeq 10.64\,\mbox{meV},\hspace{1cm}b\simeq 30.89\,\mbox{meV},\hspace{1cm}%
c\simeq -19.79\,\mbox{meV},\hspace{1cm}d\simeq (1.59+i\,5.83)\,\mbox{meV},\hspace{1cm}%
\theta \simeq 26.29^{\circ }.  \label{Parameterfit-NH}
\end{equation}
This shows that our predictive model successfully describes the current
neutrino oscillation experimental data. As $c$ depends on $b$ and $\theta$, we conclude that with only four effective
parameters, i.e., $a$, $b$, $d$ and $\theta$, we can successfully
reproduce the experimental values of the six physical observables of the
neutrino sector: the neutrino mass squared differences, the leptonic mixing
angles and the leptonic CP phase. The correlations between neutrino observables are depicted in Figure \ref{fig:correlationleptonsector}, while the value of $\theta_{23}$ is almost constant. To obtain this Figure, the lepton sector parameters were randomly generated in a range of values where the neutrino mass squared splittings, leptonic mixing parameters and leptonic CP violating phase are inside the $3\sigma$ experimentally allowed range. We note also that obtaining the correct scale for the light neutrino masses (and therefore, for the effective parameters) is implicitly setting a magnitude for $v_{123,1}/\Lambda\lesssim 10^{-2}$.

Another important lepton sector observable is the effective Majorana neutrino mass parameter of the neutrinoless double beta decay, which gives us information on the Majorana nature of neutrinos. The amplitude for this process is directly proportional to the effective Majorana mass parameter, which is defined as follows: 
\be
m_{\beta\beta}=\left\vert \sum_j U_{ek}^2m_{\nu _k}\right\vert ,  \label{mee}
\ee%
where $U_{ej}$ and $m_{\nu _k}$ are the PMNS leptonic mixing matrix
elements and the neutrino Majorana masses, respectively. 

Figure \ref{fig:mbb_vs_mminn} displays $m_{\beta\beta}$ as function of the smallest of the light active neutrino masses $m_{\rm min}$, which for the normal mass hierarchy scenario corresponds to $m_{\rm min}=m_{\nu 1}$. The points displayed are all consistent with the experimental data with a $\chi^2<1.5$. We find that our model predicts the effective Majorana neutrino mass parameter in the range  $m_{\beta\beta}\lesssim$ $(3 - 18)$ meV for the case of normal hierarchy. The new limit $T^{0\nu\beta\beta}_{1/2}({}^{100}\mathrm{Mo})\geq 1.5\times 10^{24}$ yr %for neutrinoless double-
on the half-life of $0\nu\beta\beta$ decay in $^{100}$Mo has been recently obtained \cite{Armengaud:2020luj}. This new limit translates into a corresponding upper bound on $m_{\beta\beta}\leq (300-500)$ meV at $90\%$ CL. 
However, it is worth mentioning that the proposed nEXO experiment \cite{Albert:2017hjq,Barabash:2019suz} will reach a sensitivity for the ${}^{136}\mathrm{Xe}$ $0\nu\beta\beta$ half-life of $T^{0\nu\beta\beta}_{1/2}({}^{136}\mathrm{Xe}) \geq 9.2 \times 10^{27}$ yr at $90\%$ CL. This can be converted into an exclusion limit on the effective Majorana neutrino mass between $5.7$ meV and $17.7$ meV. In the most optimistic scenario this will exclude most of the predicted region of values of our model.

%-----------------------------------------------------
\section{Constraints from FCNCs \label{sec:FCNC}}
%-----------------------------------------------------
Given the Yukawa couplings discussed in the previous Section for quarks and leptons, we have generically concluded that FCNCs will be present in both sectors, mediated by physical neutral Higgs fields. In this Section, we discuss in more detail how specific processes already act to constrain the parameter space of the model and highlight the near-future experiments that will further act to probe the model.
To this aim, we performed a numerical simulation in SPheno considering the free input parameters of our model in the following intervals
\begin{eqnarray}
&&{\rm effective\,parameters:}\quad\cfrac{v_{3}}{\Lambda} \in [0.2,0.5]\quad,\quad \cfrac{v_{i\neq 3}}{\Lambda} \in [10^{-3},0.5]\,,\\[5pt]
&&{\rm scalar\,potential:}\quad  \mu^2_h\in[0,10^6]\,{\rm GeV^2}\quad ,\quad r_{1,2}\,,\,d,\,s\,,\,\alpha_{1,2}\in [0.05,2]\quad,\quad \beta_{1,2}\in [0.05,2]\,\cfrac{v_{123}}{\Lambda}\,,\\[5pt]
 &&{\rm quark\,sector:}\quad\quad x^{(U,D)}_{ij}\,,\,c^{(U,D)}_{S_2,A}\in \pm\,[0.5,1.5] 
%  \quad \Rightarrow \quad \left( \lambda^{(U,D)}_{H^0_{1}}\in \quad,\quad \lambda^{(U,D)}_{H^0_{2}}\in\right)\,,
\,,\\[8pt]
&&{\rm neutrino\,sector:}\quad y^{(\nu)}_{1,2,3,4,5}\in[0.5,1.5]\quad,\quad m_{N_{1,2,3,4}} \in [10^{-1},10^6]\,{\rm GeV}\quad \text{with}\quad  m_{N_{1}}<m_{N_{4}}<m_{N_{2}}<m_{N_{3}}\,. \hspace{1cm}
\end{eqnarray}
\ivo{Despite the large number of model parameters, we recall that the neutrino sector of the model is predictive, as it depends only on 4 indepedent effective parameters ($a, b, d, \theta$).} \ant{Furthermore, the effective parameters $x^{(U,D)}_{ij}$ and $c^{(U,D)}_{S_2,A}$ give subleading contributions to the SM quark mass matrices and thus to the quark sector observables. Those parameters are however important as they govern the FCNCs in the quark sector.}
The most relevant Yukawa couplings are $Y^{(U,D)}_{H^0_{1,2}}$, which we analyse through the effective coupling $\lambda^{(U,D)}_{H_{1,2}}$ of Eq.~(\ref{eq:YHUD_eff}). 
Considering Eq.~(\ref{eq:YHUD_eff}), upper bounds on this quantity reflect on the ratio of the flavon VEVs and the scale $\Lambda$.
Due to the top quark Yukawa coupling, coming from Eq.~(\ref{eq:MU_MD}), we expect $v_3/\Lambda\in [0.2,0.5]$, therefore we interpret the upper bounds on $\lambda^{(U,D)}_{H^{0}_{1,2}}$ to imply a hierarchy between $v_{3}$ (larger) and $v_{123}$, $v_{1}$ (smaller). This hierarchy suppresses the effective couplings of Eq.~(\ref{eq:YHUD_eff}).
Notice that the value of these couplings is not constrained by the model itself, although there is a dependence of $v_{123,1}/\Lambda$ on the scale of light neutrino masses which suggests $\lambda^{(U,D)}_{H_{1,2}} \lesssim 5 \times 10^{-3}$.
 
\begin{figure}[p]
    %\vspace{-1.cm}
    \centering
        \hspace{5mm}\includegraphics[width=0.54\textwidth]{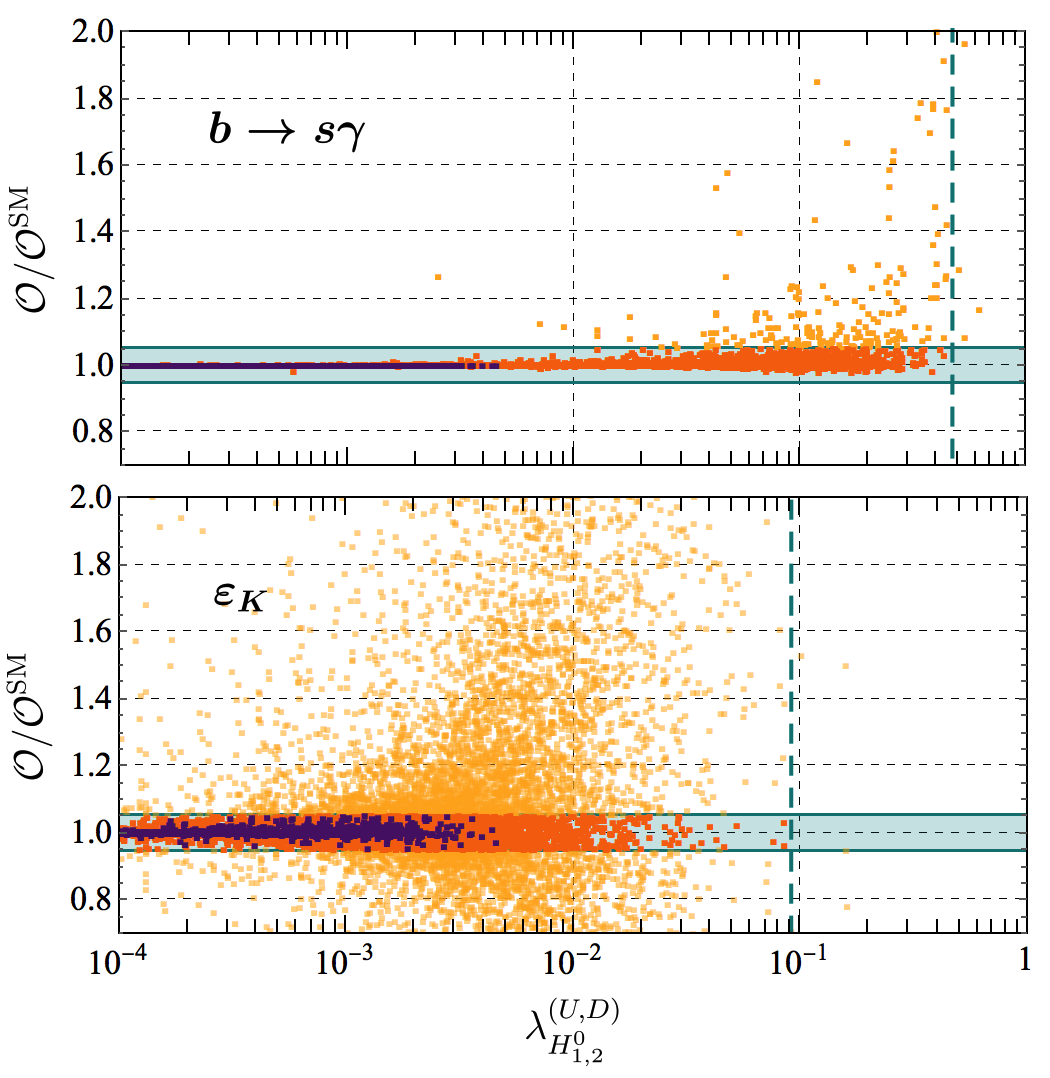}
        \captionsetup{width=0.6\textwidth%, labelsep=none
    }\caption{\label{fig:SMlimit} 
   \aur{ Limits on the effective coupling $\lambda^{(U,D)}_{H^0_{1,2}}$ (vertical dashed lines) coming from the observable $b\rightarrow s \gamma$ (top plot) and the CP observable $\varepsilon_K$ (bottom plot) respectively. The horizontal band indicates the allowed range. Light yellow points show the full scanned region. Points that satisfy the quark observable are in orange, only dark purple points are compatible with the neutrino observables.}}
        \includegraphics[width=0.56\textwidth]{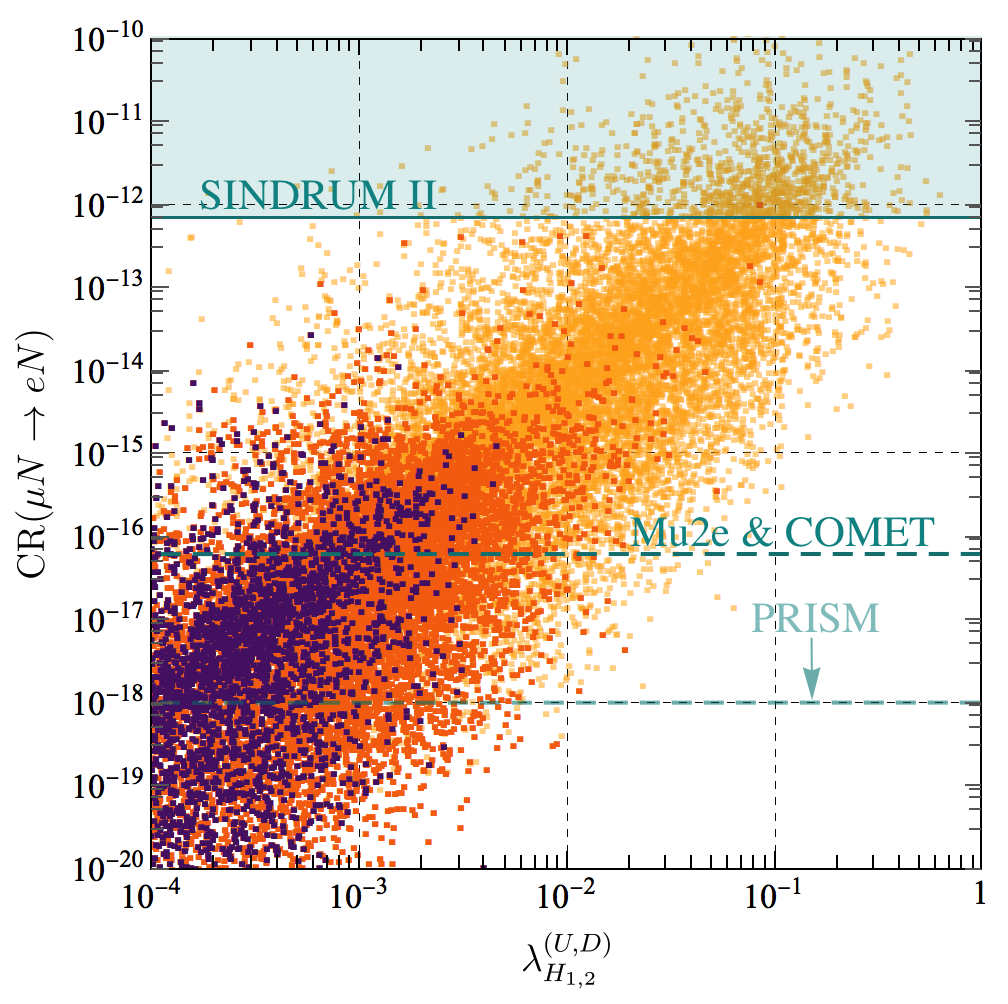}
    \captionsetup{width=0.63\textwidth %, labelsep=none
    }\caption{\label{fig:mutoe} The LFV observable $\mu \rightarrow e$ nuclear conversion versus the effective coupling $\lambda^{(U,D)}_{H^0_{1,2}}$. The colours correspond to those of \aur{ the most restrictive case of Figure \ref{fig:SMlimit} (bottom panel)}.}
    \label{fig:ModII_results}
\end{figure}

In Figure \ref{fig:SMlimit} we show the dependence of the quark flavor violating observables $b\rightarrow s\,\gamma$ and $\varepsilon_K$ on $\lambda^{(U,D)}_{H_{1,2}}$. The model prediction for these observables is displayed through ratios to their respective SM values. The narrow horizontal band indicates the limit where the experimentally allowed SM-like values are safely recovered. Figure \ref{fig:SMlimit} shows that $b\rightarrow s \gamma$ would constrain the value of the couplings to be below $\lambda^{(U,D)}_{H^{0}_{1,2}}\lesssim 0.5$ \mar{(orange points)}. 
Very similar bounds, for simplicity not displayed in the figure, come from $B_{d(s)}\rightarrow \mu^+\mu^-$ and $B_{d}\rightarrow \tau^+\tau^-$.
%%%
Figure \ref{fig:SMlimit} also shows that a more constraining limit comes from the CP violating observable $\varepsilon_K$. We see that this observable would effectively restrict the couplings to $\lambda^{(U,D)}_{H^{0}_{1,2}}\lesssim 0.1$, although most of the points concentrate at $\lambda^{(U,D)}_{H^{0}_{1,2}}\lesssim 10^{-2}$ \mar{(orange points)}.
In these plots, the orange and yellow %lighter 
points are excluded by this and other constraints, mostly the requirement to obtain the light neutrino masses.
\mar{In these plots, the requirement to obtain the light neutrino masses is very restrictive and only dark purple points satisfy all the constraints.}

In the leptonic sector, among the Lepton Flavour Violating (LFV) processes, the muon to electron flavour violating nuclear conversion
\begin{equation}
    \mu^- + N(A,Z) \rightarrow e^- + N(A,Z) 
\end{equation}
is known to provide a very sensitive probe of lepton flavour violation. Currently the best upper bound on the $\mu\rightarrow e$ nuclear conversion comes from the SINDRUM II experiment \cite{Bertl:2006up} at PSI, using a Gold stopping target. This gives a current limit on the conversion rate of CR$(\mu^-\,{\rm Au}\rightarrow e^- \,{\rm Au}) < 7 \times 10^{-13}$.

Searches for $\mu \rightarrow e$ conversion at the Mu2e experiment \cite{Bernstein:2019fyh} in FNAL and the proposed upgrade to COMET (Phase-II) experiment \cite{Kuno:2013mha} in J-PARC would achieve a similar sensitivity and an upper limit of CR$(\mu^-\,{\rm Al}\rightarrow e^-\,{\rm Al}) < 6 \times 10^{-17}$, that is four orders of magnitude below the present bound. In the long run, the PRISM/PRIME \cite{Barlow:2011zza} 
%experiment in J-PARC and the Project-X [] experiment in FNAL 
is being designed to probe values of the $\mu \rightarrow e$ conversion rate on Titanium, which is smaller by 2 orders of magnitude: CR$(\mu^-\,{\rm Ti} \rightarrow e^-\,{\rm Ti}) < 10^{-18}$.

We focus here on the $\mu \rightarrow e$ conversion because, contrary to the naive expectation of $\mu \rightarrow e$ nuclear conversion being proportional to $\mu \rightarrow e \gamma$, in our model we observe an interesting enhancement of the $\mu\rightarrow e$ nuclear conversion detached from other LFV processes like $\mu \rightarrow e\,\gamma$, $\tau \rightarrow (e,\mu)\,\gamma$, $\mu \rightarrow 3\,e$  and $\tau \rightarrow 3\,(e,\mu) $, which remain suppressed. % at the level of $\sim 10^{-40}$.
In fact, from the couplings in Eqs.~(\ref{eq:Lyl}), $\mu\rightarrow e$ can be generated already at tree-level through the exchange of a neutral scalar. % $S^0_1$.

%------------------
%------------------

Because of this, the impressive future sensitivity in this process will place significant constraints on the proposed model.

As the process also involves quarks (inside the nuclei), we find it convenient to show the observable in terms of the effective parameters $\lambda^{(U,D)}_{H^{0}_{1,2}}$ (already used in the previous Figures) 
in Figure \ref{fig:mutoe}.
The \mar{orange (grey and lightest grey)} %shaded 
points regions are already excluded by light neutrino masses, the observed value of $b\rightarrow s\,\gamma$ or $\varepsilon_K$. The dashed horizontal lines show the future limits (as discussed above).
\ivo{Particularly from Figure \ref{fig:mutoe}}, We observe that \ivo{a large percentage} of the predicted points of the model reside in a window accessible to future experiments.

\begin{figure}[t]
\centering
\includegraphics[width=0.9\textwidth]{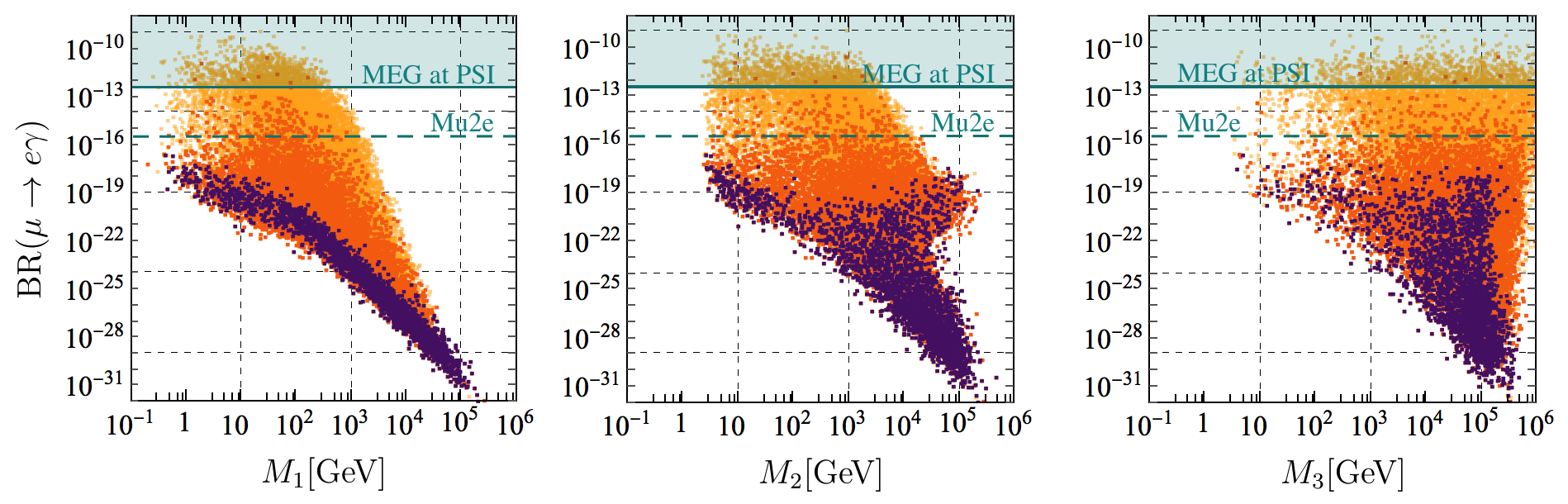}
\captionsetup{width=0.9\textwidth}
\caption{The LFV observable BR$(\mu \rightarrow e \gamma)$ versus the RH-neutrino masses. Here we have taken into account the lower bound of $[0.1-1]$ GeV on the right handed Majorana neutrino masses arising from Big Bang Nucleosynthesis (BBN) \cite{Deppisch:2015qwa}. The colours correspond to those of \aur{ the most restrictive case of Figure \ref{fig:SMlimit} (bottom panel)}.}
\label{fig:MEG_RHN}
\end{figure}

Figure \ref{fig:MEG_RHN} on the other hand shows that, while it is in theory possible to constrain the values of the RH neutrino masses through $\mu \rightarrow e \gamma$ such that it would eventually lead to lower bounds on $M_2$ and $M_3$, in practice the values expected in our model are too small to allow this process to effectively probe the parameter space.

%-----------------------------------------------------
\section{Conclusions \label{sec:conclusion}}
%-----------------------------------------------------
In this work we presented a model based on the $\Delta(27)$ family symmetry, featuring a low energy scalar potential with 3+1 $SU(2)$ doublet scalars arranged as an anti-triplet ($H)$ and trivial singlet ($h$) of the family symmetry. The latter does not acquire a Vacuum Expectation Value since it is charged under a preserved $Z_2^{\left( 1\right)}$ symmetry, and is secluded in the neutrino sector, where it leads to a radiative seesaw mechanism that produces the tiny masses of the light active neutrinos.

The quarks, being singlets of $\Delta(27)$, couple to $H$ through $\Delta(27)$ invariant combinations of $H$ and at least one of the $\Delta(27)$ triplet flavons - one such combination giving rise to their masses and mixing through the third component $H_3$ (identified with the Standard Model-like Higgs), and other combinations giving rise to Yukawa couplings to $H_{1,2}$. The extra physical scalars are mixtures of $H_{1,2}$ and have off-diagonal couplings to the quarks that are controlled by the symmetries.

The $SU(2)$ doublet leptons are arranged like $H$ as anti-triplets of $\Delta(27)$. The respective invariant combinations don't involve the triplet flavons, and include couplings to $H_3$ leading to the charged lepton masses and to $H_{1,2}$ yielding Flavour Changing Neutral Currents, which are nevertheless controlled by the symmetries.
The specific combination of neutrino masses that originate from radiative seesaw and through invariants featuring $\Delta(27)$ triplet flavons produces the cobimaximal mixing pattern.

Our model successfully accommodates the experimental values of the quark and lepton (including neutrino) masses, mixing angles, and CP phases. Furthermore, the effective Majorana neutrino mass parameter is predicted to be in the range $3$~meV$\lesssim m_{\beta\beta}\lesssim$ $18$ meV for the case of normal hierarchy. \mar{Most of the} predicted range of values for the effective Majorana neutrino mass parameter is within the declared range $5.7-17.7$ meV of sensitivity of modern experiments \cite{Barabash:2019suz}.

The detailed analysis of the Flavour Changing Neutral Currents lead to strong constraints on the model parameter space. Of particular note are $\mu \to e$  nuclear conversion processes and Kaon mixing, which already restrict the model parameter space, and that are generally predicted by the model to be in a range within the reach of future experiments.

%-----------------------------------------------------
\section*{Acknowledgments}
%-----------------------------------------------------
The authors thank Avelino Vicente for very useful discussions. A.E.C.H is supported by ANID-Chile FONDECYT 1210378. IdMV acknowledges funding from Funda\c{c}\~{a}o para a Ci\^{e}ncia e a Tecnologia (FCT) through the contract IF/00816/2015 and was supported in part by the National Science Center, Poland, through the HARMONIA project under contract UMO-2015/18/M/ST2/00518, and by FCT through projects CFTP-FCT Unit 777 (UID/FIS/00777/2019), PTDC/FIS-PAR/29436/2017, CERN/FIS-PAR/0004/2019 and CERN/FIS-PAR/0008/2019 which are partially funded through POCTI (FEDER), COMPETE, QREN and EU. 
MLLI acknowledges support from the China Postdoctoral Science Foundation No.2020M670475.
AM acknowledges support by the Estonian Research Council grants PRG803 and MOBTT86, and by the EU through the European Regional Development Fund CoE program TK133 ``The Dark Side of the Universe."

%-----------------------------------------------------
\appendix
%-----------------------------------------------------
%------------------
\section{The $\Delta(27)$ discrete group}
%------------------

The $\Delta (27)$ discrete group has the following 11 irreducible
representations: one triplet $\mathbf{3}$, one anti-triplet $\overline{%
\mathbf{3}}$ and nine singlets $\mathbf{1}_{k,l}$ ($k,l=0,1,$), where $k$
and $l$ correspond to the charges of two $Z_{3}$ and $Z_{3}^{\prime }$ generators of this group,
respectively \cite{Ishimori:2010au}. The $\Delta (27)$ irreducible representations fulfill the following tensor product rules \cite{Ishimori:2010au}:
\begin{eqnarray}
\mathbf{3}\otimes \mathbf{3} &=&\overline{\mathbf{3}}_{S_{1}}\oplus 
\overline{\mathbf{3}}_{S_{2}}\oplus \overline{\mathbf{3}}_{A}  \notag \\
\overline{\mathbf{3}}\otimes \overline{\mathbf{3}} &=&\mathbf{3}%
_{S_{1}}\oplus \mathbf{3}_{S_{2}}\oplus \mathbf{3}_{A}  \notag \\
\mathbf{3}\otimes \overline{\mathbf{3}} &=&\sum_{r=0}^{2}\mathbf{1}%
_{r,0}\oplus \sum_{r=0}^{2}\mathbf{1}_{r,1}\oplus \sum_{r=0}^{2}\mathbf{1}%
_{r,2}  \notag \\
\mathbf{1}_{k,\ell }\otimes \mathbf{1}_{k^{\prime },\ell ^{\prime }} &=&%
\mathbf{1}_{k+k^{\prime }mod3,\ell +\ell ^{\prime }mod3} \\
&& \notag
\end{eqnarray}%
Denoting $\left( x_{1},y_{1},z_{1}\right) $ and $\left(
x_{2},y_{2},z_{2}\right) $ as the basis vectors for two $\Delta (27)$ triplets $\mathbf{3}$ (or $\overline{\mathbf{3}}$), one finds: 
\begin{eqnarray}
\left( \mathbf{3}\otimes \mathbf{3}\right) _{\overline{\mathbf{3}}_{S_{1}}}
&=&\left( x_{1}y_{1},x_{2}y_{2},x_{3}y_{3}\right) ,  \notag
\label{triplet-vectors} \\
\left( \mathbf{3}\otimes \mathbf{3}\right) _{\overline{\mathbf{3}}_{S_{2}}}
&=&\frac{1}{2}\left(
x_{2}y_{3}+x_{3}y_{2},x_{3}y_{1}+x_{1}y_{3},x_{1}y_{2}+x_{2}y_{1}\right) , 
\notag \\
\left( \mathbf{3}\otimes \mathbf{3}\right) _{\overline{\mathbf{3}}_{A}} &=&%
\frac{1}{2}\left(
x_{2}y_{3}-x_{3}y_{2},x_{3}y_{1}-x_{1}y_{3},x_{1}y_{2}-x_{2}y_{1}\right) , 
\notag \\
\left( \mathbf{3}\otimes \overline{\mathbf{3}}\right) _{\mathbf{1}_{r,0}}
&=&x_{1}y_{1}+\omega ^{2r}x_{2}y_{2}+\omega ^{r}x_{3}y_{3},  \notag \\
\left( \mathbf{3}\otimes \overline{\mathbf{3}}\right) _{\mathbf{1}_{r,1}}
&=&x_{1}y_{2}+\omega ^{2r}x_{2}y_{3}+\omega ^{r}x_{3}y_{1},  \notag \\
\left( \mathbf{3}\otimes \overline{\mathbf{3}}\right) _{\mathbf{1}_{r,2}}
&=&x_{1}y_{3}+\omega ^{2r}x_{2}y_{1}+\omega ^{r}x_{3}y_{2},
\end{eqnarray}%
where $r=0,1,2$ and $\omega =e^{i\frac{2\pi }{3}}$.

%------------------
\section{Scalar potential \label{app:D27pot}}
%------------------

The scalar potential for the $\Delta(27)$ triplet $H$ can be written in the form:
\begin{eqnarray}
V_{H} &=&-\mu _{H}^{2}\left( HH^{\dagger }\right) _{\mathbf{\mathbf{1}_{0%
\mathbf{,0}}}}+\rho_{1}\left( HH^{\dagger }\right) _{\mathbf{\mathbf{1}_{0%
\mathbf{,0}}}}\left( HH^{\dagger }\right) _{\mathbf{\mathbf{1}_{0\mathbf{,0}}%
}}+\rho_{2}\left( HH^{\dagger }\right) _{\mathbf{\mathbf{1}_{1\mathbf{,0}}%
}}\left( HH^{\dagger }\right) _{\mathbf{\mathbf{1}_{2\mathbf{,0}}}}+\rho_{3}\left( HH^{\dagger }\right) _{\mathbf{\mathbf{1}_{0,1}}}\left(
HH^{\dagger }\right) _{\mathbf{\mathbf{1}_{0,2}}}  \notag \\
&&+\rho_{4}\left[ \left( HH^{\dagger }\right) _{\mathbf{\mathbf{1}_{1,1}}%
}\left( HH^{\dagger }\right) _{\mathbf{\mathbf{1}_{2,2}}}+\text{h.c.}\right].
\label{eq:VHD27inv}
\end{eqnarray}

The following relations hold between the parameters in Eq.~\eqref{eq:potV0H} and those in Eq.~\eqref{eq:VHD27inv}:
\begin{equation}
    s \;\equiv\; \rho_1 + \rho_2 \qquad,\qquad
    r_1 \;\equiv\; 2\rho_1 - \rho_2 \qquad,\qquad
    r_2 \;\equiv\; \rho_3 - \rho_4 \qquad,\qquad
    d \;\equiv\; \rho_3 - \omega^2 \rho_4 \,.
\end{equation}

%------------------
\section{Neutrino mass parameters\label{app:Mnupar}}
%------------------
The neutrino Yukawa matrix, in the basis of diagonal RH-neutrinos, is given by
\begin{equation}
    \widetilde{Y}_\nu ~=~ Y_\nu\, R_\beta^T ~=~ \frac{1}{\sqrt{2}\, \Lambda}
    \begin{pmatrix}
        z_1^{(\nu)} & z_3^{(\nu)} & 0 \\
        z_2^{(\nu)} \omega^2 & z_4^{(\nu)} \omega^2 & ~~z_5^{(\nu)} \\
        z_2^{(\nu)} \omega & z_4^{(\nu)} \omega & -z_5^{(\nu)}
    \end{pmatrix} ,
\end{equation}
where $R_\beta$ refers to the rotation in Eq.\eqref{eq:N12tilde} and the Yukawa parameters are:
\begin{eqnarray}
\label{eq:znui}
z_{1}^{\left( \nu \right) } &=&\frac{1}{\sqrt{2}}\left[ \left( y_{3}^{\left( \nu \right) }\frac{v_{123}%
}{\Lambda }+y_{1}^{\left(
\nu \right) }\frac{v_{1}}{\Lambda }\right) \cos \beta -\left( y_{2}^{\left( \nu \right) }\frac{%
v_{123}}{\Lambda }+y_{4}^{\left( \nu \right) }\frac{v_{1}}{\Lambda }\right)
\sin \beta \right] ,\notag \\
z_{2}^{\left( \nu \right) } &=&\frac{v_{123}}{\sqrt{2}\Lambda }\left(
y_{3}^{\left( \nu \right) }\cos \beta -y_{2}^{\left( \nu \right) }\sin
\beta \right) , \notag\\
z_{3}^{\left( \nu \right) } &=&\frac{1}{\sqrt{2}}\left[ \left( y_{3}^{\left( \nu \right) }\frac{v_{123}%
}{\Lambda } +y_{1}^{\left(
\nu \right) }\frac{v_{1}}{\Lambda }\right) \sin \beta +\left( y_{2}^{\left( \nu \right) }\frac{%
v_{123}}{\Lambda }+y_{4}^{\left( \nu \right) }\frac{v_{1}}{\Lambda }\right)
\cos \beta \right] , \\
z_{4}^{\left( \nu \right) } &=&\frac{v_{123}}{\sqrt{2}\Lambda }\left(
y_{3}^{\left( \nu \right) }\sin \beta +y_{2}^{\left( \nu \right) }\cos
\beta \right), \notag\\
z_{5}^{\left( \nu \right) } &=&\frac{y_{5}^{\left( \nu \right) }v_{23}}{%
\sqrt{2}\Lambda }.\notag
\end{eqnarray}
The relation between the light effective neutrino mass parameters in Eq.~(\ref{eq:Mnu}) and the lagrangian parameters $z^{\nu}_i$ reads as 
\begin{eqnarray}
a &=&\frac{\left( z_{1}^{\left( \nu \right) }\right) ^{2}m_{\widetilde{N}_{1}}}{16\pi
^{2}}f_{1}+\frac{\left( z_{3}^{\left( \nu \right) }\right) ^{2}m_{\widetilde{N}_{2}}}{%
16\pi ^{2}}f_{2},  \notag \\
b &=&\left\vert \frac{\omega \left( z_{2}^{\left( \nu \right) }\right)
^{2}m_{\widetilde{N}_{1}}}{16\pi ^{2}}f_{1}+\frac{\omega \left( z_{4}^{\left( \nu
\right) }\right) ^{2}m_{\widetilde{N}_{2}}}{16\pi ^{2}}f_{2}+\frac{\left( z_{5}^{\left(
\nu \right) }\right) ^{2}m_{N_{3}}}{16\pi ^{2}}f_{3}\right\vert ,  \notag \\
c &=&\frac{\left( z_{2}^{\left( \nu \right) }\right) ^{2}m_{\widetilde{N}_{1}}}{16\pi
^{2}}f_{1}+\frac{\left( z_{4}^{\left( \nu \right) }\right) ^{2}m_{\widetilde{N}_{2}}}{%
16\pi ^{2}}f_{2}-\frac{\left( z_{5}^{\left( \nu \right) }\right)
^{2}m_{N_{3}}}{16\pi ^{2}}f_{3} , \\
d &=&\frac{z_{1}^{\left( \nu \right) }z_{2}^{\left( \nu \right) }m_{\widetilde{N}_{1}}}{%
16\pi ^{2}}f_{1}+\frac{z_{3}^{\left( \nu \right) }z_{4}^{\left( \nu \right)
}m_{\widetilde{N}_{2}}}{16\pi ^{2}}f_{2},  \notag \\
\theta  &=&\arg \left( \frac{\omega \left( z_{2}^{\left( \nu \right)
}\right) ^{2}m_{\widetilde{N}_{1}}}{16\pi ^{2}}f_{1}+\frac{\omega \left( z_{4}^{\left(
\nu \right) }\right) ^{2}m_{\widetilde{N}_{2}}}{16\pi ^{2}}f_{2}+\frac{\left(
z_{5}^{\left( \nu \right) }\right) ^{2}m_{N_{3}}}{16\pi ^{2}}f_{3}\right) ,\notag
\end{eqnarray}
\vspace{-5mm}
with $f_k$ as defined in Eq.~(\ref{eq:fk}). The system admits a solution as long as $c=b\,(\sin\theta -\sqrt{3}\cos\theta)/\sqrt{3}$.


\begin{thebibliography}{10}

\bibitem{Branco:1983tn}
G.~C. Branco, J.~M. Gerard, and W.~Grimus, ``{GEOMETRICAL T VIOLATION},''
\href{http://dx.doi.org/10.1016/0370-2693(84)92024-0}{{\em Phys. Lett.}
  {\bfseries 136B} (1984) 383--386}.
%%CITATION = PHLTA,136B,383;%%.

\bibitem{deMedeirosVarzielas:2006fc}
I.~de~Medeiros~Varzielas, S.~F. King, and G.~G. Ross, ``{Neutrino
  tri-bi-maximal mixing from a non-Abelian discrete family symmetry},''
  \href{http://dx.doi.org/10.1016/j.physletb.2007.03.009}{{\em Phys. Lett.}
  {\bfseries B648} (2007) 201--206},
\href{http://arxiv.org/abs/hep-ph/0607045}{{\ttfamily arXiv:hep-ph/0607045
  [hep-ph]}}.
%%CITATION = HEP-PH/0607045;%%.

\bibitem{Ma:2006ip}
E.~Ma, ``{Neutrino Mass Matrix from Delta(27) Symmetry},''
  \href{http://dx.doi.org/10.1142/S0217732306021190}{{\em Mod. Phys. Lett.}
  {\bfseries A21} (2006) 1917--1921},
\href{http://arxiv.org/abs/hep-ph/0607056}{{\ttfamily arXiv:hep-ph/0607056
  [hep-ph]}}.
%%CITATION = HEP-PH/0607056;%%.

\bibitem{Ma:2007wu}
E.~Ma, ``{Near tribimaximal neutrino mixing with Delta(27) symmetry},''
  \href{http://dx.doi.org/10.1016/j.physletb.2007.12.060}{{\em Phys. Lett.}
  {\bfseries B660} (2008) 505--507},
\href{http://arxiv.org/abs/0709.0507}{{\ttfamily arXiv:0709.0507 [hep-ph]}}.
%%CITATION = ARXIV:0709.0507;%%.

\bibitem{Bazzocchi:2009qg}
F.~Bazzocchi and I.~de~Medeiros~Varzielas, ``{Tri-bi-maximal mixing in viable
  family symmetry unified model with extended seesaw},''
  \href{http://dx.doi.org/10.1103/PhysRevD.79.093001}{{\em Phys. Rev.}
  {\bfseries D79} (2009) 093001},
\href{http://arxiv.org/abs/0902.3250}{{\ttfamily arXiv:0902.3250 [hep-ph]}}.
%%CITATION = ARXIV:0902.3250;%%.

\bibitem{deMedeirosVarzielas:2011zw}
I.~de~Medeiros~Varzielas and D.~Emmanuel-Costa, ``{Geometrical CP Violation},''
  \href{http://dx.doi.org/10.1103/PhysRevD.84.117901}{{\em Phys. Rev.}
  {\bfseries D84} (2011) 117901},
\href{http://arxiv.org/abs/1106.5477}{{\ttfamily arXiv:1106.5477 [hep-ph]}}.
%%CITATION = ARXIV:1106.5477;%%.

\bibitem{Varzielas:2012nn}
I.~de~Medeiros~Varzielas, D.~Emmanuel-Costa, and P.~Leser, ``{Geometrical CP
  Violation from Non-Renormalisable Scalar Potentials},''
  \href{http://dx.doi.org/10.1016/j.physletb.2012.08.008}{{\em Phys. Lett.}
  {\bfseries B716} (2012) 193--196},
\href{http://arxiv.org/abs/1204.3633}{{\ttfamily arXiv:1204.3633 [hep-ph]}}.
%%CITATION = ARXIV:1204.3633;%%.

\bibitem{Bhattacharyya:2012pi}
G.~Bhattacharyya, I.~de~Medeiros~Varzielas, and P.~Leser, ``{A common origin of
  fermion mixing and geometrical CP violation, and its test through Higgs
  physics at the LHC},''
  \href{http://dx.doi.org/10.1103/PhysRevLett.109.241603}{{\em Phys. Rev.
  Lett.} {\bfseries 109} (2012) 241603},
\href{http://arxiv.org/abs/1210.0545}{{\ttfamily arXiv:1210.0545 [hep-ph]}}.
%%CITATION = ARXIV:1210.0545;%%.

\bibitem{Ferreira:2012ri}
P.~M. Ferreira, W.~Grimus, L.~Lavoura, and P.~O. Ludl, ``{Maximal CP Violation
  in Lepton Mixing from a Model with Delta(27) flavour Symmetry},''
  \href{http://dx.doi.org/10.1007/JHEP09(2012)128}{{\em JHEP} {\bfseries 09}
  (2012) 128},
\href{http://arxiv.org/abs/1206.7072}{{\ttfamily arXiv:1206.7072 [hep-ph]}}.
%%CITATION = ARXIV:1206.7072;%%.

\bibitem{Ma:2013xqa}
E.~Ma, ``{Neutrino Mixing and Geometric CP Violation with Delta(27)
  Symmetry},'' \href{http://dx.doi.org/10.1016/j.physletb.2013.05.011}{{\em
  Phys. Lett.} {\bfseries B723} (2013) 161--163},
\href{http://arxiv.org/abs/1304.1603}{{\ttfamily arXiv:1304.1603 [hep-ph]}}.
%%CITATION = ARXIV:1304.1603;%%.

\bibitem{Nishi:2013jqa}
C.~C. Nishi, ``{Generalized $CP$ symmetries in $\Delta(27)$ flavor models},''
  \href{http://dx.doi.org/10.1103/PhysRevD.88.033010}{{\em Phys. Rev.}
  {\bfseries D88} no.~3, (2013) 033010},
\href{http://arxiv.org/abs/1306.0877}{{\ttfamily arXiv:1306.0877 [hep-ph]}}.
%%CITATION = ARXIV:1306.0877;%%.

\bibitem{Varzielas:2013sla}
I.~de~Medeiros~Varzielas and D.~Pidt, ``{Towards realistic models of quark
  masses with geometrical CP violation},''
  \href{http://dx.doi.org/10.1088/0954-3899/41/2/025004}{{\em J. Phys.}
  {\bfseries G41} (2014) 025004},
\href{http://arxiv.org/abs/1307.0711}{{\ttfamily arXiv:1307.0711 [hep-ph]}}.
%%CITATION = ARXIV:1307.0711;%%.

\bibitem{Aranda:2013gga}
A.~Aranda, C.~Bonilla, S.~Morisi, E.~Peinado, and J.~W.~F. Valle, ``{Dirac
  neutrinos from flavor symmetry},''
  \href{http://dx.doi.org/10.1103/PhysRevD.89.033001}{{\em Phys. Rev.}
  {\bfseries D89} no.~3, (2014) 033001},
\href{http://arxiv.org/abs/1307.3553}{{\ttfamily arXiv:1307.3553 [hep-ph]}}.
%%CITATION = ARXIV:1307.3553;%%.

\bibitem{Varzielas:2013eta}
I.~de~Medeiros~Varzielas and D.~Pidt, ``{Geometrical CP violation with a
  complete fermion sector},''
  \href{http://dx.doi.org/10.1007/JHEP11(2013)206}{{\em JHEP} {\bfseries 11}
  (2013) 206},
\href{http://arxiv.org/abs/1307.6545}{{\ttfamily arXiv:1307.6545 [hep-ph]}}.
%%CITATION = ARXIV:1307.6545;%%.

\bibitem{Harrison:2014jqa}
P.~F. Harrison, R.~Krishnan, and W.~G. Scott, ``{Deviations from tribimaximal
  neutrino mixing using a model with $\Delta(27)$ symmetry},''
  \href{http://dx.doi.org/10.1142/S0217751X1450095X}{{\em Int. J. Mod. Phys.}
  {\bfseries A29} no.~18, (2014) 1450095},
\href{http://arxiv.org/abs/1406.2025}{{\ttfamily arXiv:1406.2025 [hep-ph]}}.
%%CITATION = ARXIV:1406.2025;%%.

\bibitem{Ma:2014eka}
E.~Ma and A.~Natale, ``{Scotogenic $Z_2$ or $U(1)_D$ Model of Neutrino Mass
  with $\Delta(27)$ Symmetry},''
  \href{http://dx.doi.org/10.1016/j.physletb.2014.05.070}{{\em Phys. Lett.}
  {\bfseries B734} (2014) 403--405},
\href{http://arxiv.org/abs/1403.6772}{{\ttfamily arXiv:1403.6772 [hep-ph]}}.
%%CITATION = ARXIV:1403.6772;%%.

\bibitem{Abbas:2014ewa}
M.~Abbas and S.~Khalil, ``{Fermion masses and mixing in $\Delta(27)$ flavour
  model},'' \href{http://dx.doi.org/10.1103/PhysRevD.91.053003}{{\em Phys.
  Rev.} {\bfseries D91} no.~5, (2015) 053003},
\href{http://arxiv.org/abs/1406.6716}{{\ttfamily arXiv:1406.6716 [hep-ph]}}.
%%CITATION = ARXIV:1406.6716;%%.

\bibitem{Abbas:2015zna}
M.~Abbas, S.~Khalil, A.~Rashed, and A.~Sil, ``{Neutrino masses and deviation
  from tribimaximal mixing in $\Delta(27)$ model with inverse seesaw
  mechanism},'' \href{http://dx.doi.org/10.1103/PhysRevD.93.013018}{{\em Phys.
  Rev.} {\bfseries D93} no.~1, (2016) 013018},
\href{http://arxiv.org/abs/1508.03727}{{\ttfamily arXiv:1508.03727 [hep-ph]}}.
%%CITATION = ARXIV:1508.03727;%%.

\bibitem{Varzielas:2015aua}
I.~de~Medeiros~Varzielas, ``{$\Delta(27)$ family symmetry and neutrino
  mixing},'' \href{http://dx.doi.org/10.1007/JHEP08(2015)157}{{\em JHEP}
  {\bfseries 08} (2015) 157},
\href{http://arxiv.org/abs/1507.00338}{{\ttfamily arXiv:1507.00338 [hep-ph]}}.
%%CITATION = ARXIV:1507.00338;%%.

\bibitem{Bjorkeroth:2015uou}
F.~Björkeroth, F.~J. de~Anda, I.~de~Medeiros~Varzielas, and S.~F. King,
  ``{Towards a complete $\Delta(27) \times SO(10)$ SUSY GUT},''
  \href{http://dx.doi.org/10.1103/PhysRevD.94.016006}{{\em Phys. Rev.}
  {\bfseries D94} no.~1, (2016) 016006},
\href{http://arxiv.org/abs/1512.00850}{{\ttfamily arXiv:1512.00850 [hep-ph]}}.
%%CITATION = ARXIV:1512.00850;%%.

\bibitem{Chen:2015jta}
P.~Chen, G.-J. Ding, A.~D. Rojas, C.~A. Vaquera-Araujo, and J.~W.~F. Valle,
  ``{Warped flavor symmetry predictions for neutrino physics},''
  \href{http://dx.doi.org/10.1007/JHEP01(2016)007}{{\em JHEP} {\bfseries 01}
  (2016) 007},
\href{http://arxiv.org/abs/1509.06683}{{\ttfamily arXiv:1509.06683 [hep-ph]}}.
%%CITATION = ARXIV:1509.06683;%%.

\bibitem{Vien:2016tmh}
V.~V. Vien, A.~E. Cárcamo~Hernández, and H.~N. Long, ``{The $\Delta(27)$
  flavor 3-3-1 model with neutral leptons},''
  \href{http://dx.doi.org/10.1016/j.nuclphysb.2016.10.010}{{\em Nucl. Phys.}
  {\bfseries B913} (2016) 792--814},
\href{http://arxiv.org/abs/1601.03300}{{\ttfamily arXiv:1601.03300 [hep-ph]}}.
%%CITATION = ARXIV:1601.03300;%%.

\bibitem{Hernandez:2016eod}
A.~E. Cárcamo~Hernández, H.~N. Long, and V.~V. Vien, ``{A 3-3-1 model with
  right-handed neutrinos based on the $\varDelta \left( 27\right) $ family
  symmetry},'' \href{http://dx.doi.org/10.1140/epjc/s10052-016-4074-0}{{\em
  Eur. Phys. J.} {\bfseries C76} no.~5, (2016) 242},
\href{http://arxiv.org/abs/1601.05062}{{\ttfamily arXiv:1601.05062 [hep-ph]}}.
%%CITATION = ARXIV:1601.05062;%%.

\bibitem{Bjorkeroth:2016lzs}
F.~Björkeroth, F.~J. de~Anda, I.~de~Medeiros~Varzielas, and S.~F. King,
  ``{Leptogenesis in a $\Delta(27) \times SO(10)$ SUSY GUT},''
  \href{http://dx.doi.org/10.1007/JHEP01(2017)077}{{\em JHEP} {\bfseries 01}
  (2017) 077},
\href{http://arxiv.org/abs/1609.05837}{{\ttfamily arXiv:1609.05837 [hep-ph]}}.
%%CITATION = ARXIV:1609.05837;%%.

\bibitem{CarcamoHernandez:2017owh}
A.~E. Cárcamo~Hernández, S.~Kovalenko, J.~W.~F. Valle, and C.~A.
  Vaquera-Araujo, ``{Predictive Pati-Salam theory of fermion masses and
  mixing},'' \href{http://dx.doi.org/10.1007/JHEP07(2017)118}{{\em JHEP}
  {\bfseries 07} (2017) 118},
\href{http://arxiv.org/abs/1705.06320}{{\ttfamily arXiv:1705.06320 [hep-ph]}}.
%%CITATION = ARXIV:1705.06320;%%.

\bibitem{deMedeirosVarzielas:2017sdv}
I.~de~Medeiros~Varzielas, G.~G. Ross, and J.~Talbert, ``{A Unified Model of
  Quarks and Leptons with a Universal Texture Zero},''
  \href{http://dx.doi.org/10.1007/JHEP03(2018)007}{{\em JHEP} {\bfseries 03}
  (2018) 007},
\href{http://arxiv.org/abs/1710.01741}{{\ttfamily arXiv:1710.01741 [hep-ph]}}.
%%CITATION = ARXIV:1710.01741;%%.

\bibitem{Bernal:2017xat}
N.~Bernal, A.~E. Cárcamo~Hernández, I.~de~Medeiros~Varzielas, and
  S.~Kovalenko, ``{Fermion masses and mixings and dark matter constraints in a
  model with radiative seesaw mechanism},''
  \href{http://dx.doi.org/10.1007/JHEP05(2018)053}{{\em JHEP} {\bfseries 05}
  (2018) 053},
\href{http://arxiv.org/abs/1712.02792}{{\ttfamily arXiv:1712.02792 [hep-ph]}}.
%%CITATION = ARXIV:1712.02792;%%.

\bibitem{CarcamoHernandez:2018iel}
A.~E. Cárcamo~Hernández, H.~N. Long, and V.~V. Vien, ``{The first
  $\Delta(27)$ flavor 3-3-1 model with low scale seesaw mechanism},''
  \href{http://dx.doi.org/10.1140/epjc/s10052-018-6284-0}{{\em Eur. Phys. J.}
  {\bfseries C78} no.~10, (2018) 804},
\href{http://arxiv.org/abs/1803.01636}{{\ttfamily arXiv:1803.01636 [hep-ph]}}.
%%CITATION = ARXIV:1803.01636;%%.

\bibitem{deMedeirosVarzielas:2018vab}
I.~De~Medeiros~Varzielas, M.~L. López-Ibáñez, A.~Melis, and O.~Vives,
  ``{Controlled flavor violation in the MSSM from a unified $\Delta(27)$ flavor
  symmetry},'' \href{http://dx.doi.org/10.1007/JHEP09(2018)047}{{\em JHEP}
  {\bfseries 09} (2018) 047},
\href{http://arxiv.org/abs/1807.00860}{{\ttfamily arXiv:1807.00860 [hep-ph]}}.
%%CITATION = ARXIV:1807.00860;%%.

\bibitem{CarcamoHernandez:2018hst}
A.~E. Cárcamo~Hernández, S.~Kovalenko, J.~W.~F. Valle, and C.~A.
  Vaquera-Araujo, ``{Neutrino predictions from a left-right symmetric flavored
  extension of the standard model},''
  \href{http://dx.doi.org/10.1007/JHEP02(2019)065}{{\em JHEP} {\bfseries 02}
  (2019) 065},
\href{http://arxiv.org/abs/1811.03018}{{\ttfamily arXiv:1811.03018 [hep-ph]}}.
%%CITATION = ARXIV:1811.03018;%%.

\bibitem{CarcamoHernandez:2018djj}
A.~E. Cárcamo~Hernández, J.~C. Gómez-Izquierdo, S.~Kovalenko, and
  M.~Mondragón, ``{$\Delta \left( 27\right)$ flavor singlet-triplet Higgs
  model for fermion masses and mixings},''
  \href{http://dx.doi.org/10.1016/j.nuclphysb.2019.114688}{{\em Nucl. Phys.}
  {\bfseries B946} (2019) 114688},
\href{http://arxiv.org/abs/1810.01764}{{\ttfamily arXiv:1810.01764 [hep-ph]}}.
%%CITATION = ARXIV:1810.01764;%%.

\bibitem{Ma:2019iwj}
E.~Ma, ``{Scotogenic cobimaximal Dirac neutrino mixing from $\Delta (27)$ and
  $U(1)_\chi $},'' \href{http://dx.doi.org/10.1140/epjc/s10052-019-7440-x}{{\em
  Eur. Phys. J.} {\bfseries C79} no.~11, (2019) 903},
\href{http://arxiv.org/abs/1905.01535}{{\ttfamily arXiv:1905.01535 [hep-ph]}}.
%%CITATION = ARXIV:1905.01535;%%.

\bibitem{Bjorkeroth:2019csz}
F.~Björkeroth, I.~de~Medeiros~Varzielas, M.~L. López-Ibáñez, A.~Melis, and
  O.~Vives, ``{Leptogenesis in $\Delta(27)$ with a Universal Texture Zero},''
  \href{http://dx.doi.org/10.1007/JHEP09(2019)050}{{\em JHEP} {\bfseries 09}
  (2019) 050},
\href{http://arxiv.org/abs/1904.10545}{{\ttfamily arXiv:1904.10545 [hep-ph]}}.
%%CITATION = ARXIV:1904.10545;%%.

\bibitem{CarcamoHernandez:2020udg}
A.~E. Cárcamo~Hernández and I.~de~Medeiros~Varzielas, ``{$\Delta(27)$
  framework for cobimaximal neutrino mixing models},''
  \href{http://dx.doi.org/10.1016/j.physletb.2020.135491}{{\em Phys. Lett.}
  {\bfseries B806} (2020) 135491},
\href{http://arxiv.org/abs/2003.01134}{{\ttfamily arXiv:2003.01134 [hep-ph]}}.
%%CITATION = ARXIV:2003.01134;%%.

\bibitem{Diaz:2015pyv}
M.~A. Díaz, B.~Koch, and S.~Urrutia-Quiroga, ``{Constraints to Dark Matter
  from Inert Higgs Doublet Model},''
  \href{http://dx.doi.org/10.1155/2016/8278375}{{\em Adv. High Energy Phys.}
  {\bfseries 2016} (2016) 8278375},
\href{http://arxiv.org/abs/1511.04429}{{\ttfamily arXiv:1511.04429 [hep-ph]}}.
%%CITATION = ARXIV:1511.04429;%%.

\bibitem{Escudero:2016gzx}
M.~Escudero, A.~Berlin, D.~Hooper, and M.-X. Lin, ``{Toward (Finally!) Ruling
  Out Z and Higgs Mediated Dark Matter Models},''
  \href{http://dx.doi.org/10.1088/1475-7516/2016/12/029}{{\em JCAP} {\bfseries
  1612} (2016) 029},
\href{http://arxiv.org/abs/1609.09079}{{\ttfamily arXiv:1609.09079 [hep-ph]}}.
%%CITATION = ARXIV:1609.09079;%%.

\bibitem{Arbelaez:2016mhg}
C.~Arbeláez, A.~E. Cárcamo~Hernández, S.~Kovalenko, and I.~Schmidt,
  ``{Radiative Seesaw-type Mechanism of Fermion Masses and Non-trivial Quark
  Mixing},'' \href{http://dx.doi.org/10.1140/epjc/s10052-017-4948-9}{{\em Eur.
  Phys. J.} {\bfseries C77} no.~6, (2017) 422},
\href{http://arxiv.org/abs/1602.03607}{{\ttfamily arXiv:1602.03607 [hep-ph]}}.
%%CITATION = ARXIV:1602.03607;%%.

\bibitem{Garcia-Cely:2015khw}
C.~Garcia-Cely, M.~Gustafsson, and A.~Ibarra, ``{Probing the Inert Doublet Dark
  Matter Model with Cherenkov Telescopes},''
  \href{http://dx.doi.org/10.1088/1475-7516/2016/02/043}{{\em JCAP} {\bfseries
  1602} (2016) 043},
\href{http://arxiv.org/abs/1512.02801}{{\ttfamily arXiv:1512.02801 [hep-ph]}}.
%%CITATION = ARXIV:1512.02801;%%.

\bibitem{Rojas-Abatte:2017hqm}
F.~Rojas-Abatte, M.~L. Mora, J.~Urbina, and A.~R. Zerwekh, ``{Inert
  two-Higgs-doublet model strongly coupled to a non-Abelian vector
  resonance},'' \href{http://dx.doi.org/10.1103/PhysRevD.96.095025}{{\em Phys.
  Rev.} {\bfseries D96} no.~9, (2017) 095025},
\href{http://arxiv.org/abs/1707.04543}{{\ttfamily arXiv:1707.04543 [hep-ph]}}.
%%CITATION = ARXIV:1707.04543;%%.

\bibitem{Dutta:2017lny}
B.~Dutta, G.~Palacio, J.~D. Ruiz-Alvarez, and D.~Restrepo, ``{Vector Boson
  Fusion in the Inert Doublet Model},''
  \href{http://dx.doi.org/10.1103/PhysRevD.97.055045}{{\em Phys. Rev.}
  {\bfseries D97} no.~5, (2018) 055045},
\href{http://arxiv.org/abs/1709.09796}{{\ttfamily arXiv:1709.09796 [hep-ph]}}.
%%CITATION = ARXIV:1709.09796;%%.

\bibitem{Nomura:2017kih}
T.~Nomura and H.~Okada, ``{A radiative seesaw model with higher order terms
  under an alternative $U(1)_{B-L}$},''
  \href{http://dx.doi.org/10.1016/j.physletb.2018.04.034}{{\em Phys. Lett.}
  {\bfseries B781} (2018) 561--567},
\href{http://arxiv.org/abs/1711.05115}{{\ttfamily arXiv:1711.05115 [hep-ph]}}.
%%CITATION = ARXIV:1711.05115;%%.

\bibitem{CarcamoHernandez:2017kra}
A.~E. Cárcamo~Hernández and H.~N. Long, ``{A highly predictive $A_{4}$
  flavour 3-3-1 model with radiative inverse seesaw mechanism},''
  \href{http://dx.doi.org/10.1088/1361-6471/aaace7}{{\em J. Phys.} {\bfseries
  G45} no.~4, (2018) 045001},
\href{http://arxiv.org/abs/1705.05246}{{\ttfamily arXiv:1705.05246 [hep-ph]}}.
%%CITATION = ARXIV:1705.05246;%%.

\bibitem{CarcamoHernandez:2017cwi}
A.~E. Cárcamo~Hernández, S.~Kovalenko, H.~N. Long, and I.~Schmidt, ``{A
  variant of 3-3-1 model for the generation of the SM fermion mass and mixing
  pattern},'' \href{http://dx.doi.org/10.1007/JHEP07(2018)144}{{\em JHEP}
  {\bfseries 07} (2018) 144},
\href{http://arxiv.org/abs/1705.09169}{{\ttfamily arXiv:1705.09169 [hep-ph]}}.
%%CITATION = ARXIV:1705.09169;%%.

\bibitem{Gao:2018xld}
C.~Gao, M.~A. Luty, and N.~A. Neill, ``{Almost Inert Higgs Bosons at the
  LHC},'' \href{http://dx.doi.org/10.1007/JHEP09(2019)043}{{\em JHEP}
  {\bfseries 09} (2019) 043},
\href{http://arxiv.org/abs/1812.08179}{{\ttfamily arXiv:1812.08179 [hep-ph]}}.
%%CITATION = ARXIV:1812.08179;%%.

\bibitem{Long:2018dun}
H.~N. Long, N.~V. Hop, L.~T. Hue, N.~H. Thao, and A.~E. Cárcamo~Hernández,
  ``{Some phenomenological aspects of the 3-3-1 model with the
  Cárcamo-Kovalenko-Schmidt mechanism},''
  \href{http://dx.doi.org/10.1103/PhysRevD.100.015004}{{\em Phys. Rev.}
  {\bfseries D100} no.~1, (2019) 015004},
\href{http://arxiv.org/abs/1810.00605}{{\ttfamily arXiv:1810.00605 [hep-ph]}}.
%%CITATION = ARXIV:1810.00605;%%.

\bibitem{CarcamoHernandez:2019cbd}
A.~E. Cárcamo~Hernández, S.~Kovalenko, R.~Pasechnik, and I.~Schmidt,
  ``{Sequentially loop-generated quark and lepton mass hierarchies in an
  extended Inert Higgs Doublet model},''
  \href{http://dx.doi.org/10.1007/JHEP06(2019)056}{{\em JHEP} {\bfseries 06}
  (2019) 056},
\href{http://arxiv.org/abs/1901.02764}{{\ttfamily arXiv:1901.02764 [hep-ph]}}.
%%CITATION = ARXIV:1901.02764;%%.

\bibitem{Bhattacharya:2019fgs}
S.~Bhattacharya, P.~Ghosh, A.~K. Saha, and A.~Sil, ``{Two component dark matter
  with inert Higgs doublet: neutrino mass, high scale validity and collider
  searches},'' \href{http://dx.doi.org/10.1007/JHEP03(2020)090}{{\em JHEP}
  {\bfseries 03} (2020) 090},
\href{http://arxiv.org/abs/1905.12583}{{\ttfamily arXiv:1905.12583 [hep-ph]}}.
%%CITATION = ARXIV:1905.12583;%%.

\bibitem{Han:2019lux}
Z.-L. Han and W.~Wang, ``{Predictive Scotogenic Model with Flavor Dependent
  Symmetry},'' \href{http://dx.doi.org/10.1140/epjc/s10052-019-7033-8}{{\em
  Eur. Phys. J.} {\bfseries C79} no.~6, (2019) 522},
\href{http://arxiv.org/abs/1901.07798}{{\ttfamily arXiv:1901.07798 [hep-ph]}}.
%%CITATION = ARXIV:1901.07798;%%.

\bibitem{CarcamoHernandez:2019xkb}
A.~E. Cárcamo~Hernández, S.~Kovalenko, R.~Pasechnik, and I.~Schmidt,
  ``{Phenomenology of an extended IDM with loop-generated fermion mass
  hierarchies},'' \href{http://dx.doi.org/10.1140/epjc/s10052-019-7101-0}{{\em
  Eur. Phys. J.} {\bfseries C79} no.~7, (2019) 610},
\href{http://arxiv.org/abs/1901.09552}{{\ttfamily arXiv:1901.09552 [hep-ph]}}.
%%CITATION = ARXIV:1901.09552;%%.

\bibitem{CarcamoHernandez:2019lhv}
A.~E. Cárcamo~Hernández, D.~T. Huong, and H.~N. Long, ``{Minimal model for
  the fermion flavor structure, mass hierarchy, dark matter, leptogenesis, and
  the electron and muon anomalous magnetic moments},''
  \href{http://dx.doi.org/10.1103/PhysRevD.102.055002}{{\em Phys. Rev.}
  {\bfseries D102} no.~5, (2020) 055002},
\href{http://arxiv.org/abs/1910.12877}{{\ttfamily arXiv:1910.12877 [hep-ph]}}.
%%CITATION = ARXIV:1910.12877;%%.

\bibitem{CarcamoHernandez:2020ehn}
A.~E. Cárcamo~Hernández, J.~W.~F. Valle, and C.~A. Vaquera-Araujo, ``{Simple
  theory for scotogenic dark matter with residual matter-parity},''
  \href{http://dx.doi.org/10.1016/j.physletb.2020.135757}{{\em Phys. Lett.}
  {\bfseries B809} (2020) 135757},
\href{http://arxiv.org/abs/2006.06009}{{\ttfamily arXiv:2006.06009 [hep-ph]}}.
%%CITATION = ARXIV:2006.06009;%%.

\bibitem{Felipe:2013vwa}
R.~Gonzalez~Felipe, H.~Serodio, and J.~P. Silva, ``{Neutrino masses and mixing
  in A4 models with three Higgs doublets},''
  \href{http://dx.doi.org/10.1103/PhysRevD.88.015015}{{\em Phys. Rev.}
  {\bfseries D88} no.~1, (2013) 015015},
\href{http://arxiv.org/abs/1304.3468}{{\ttfamily arXiv:1304.3468 [hep-ph]}}.
%%CITATION = ARXIV:1304.3468;%%.

\bibitem{Fukuura:1999ze}
K.~Fukuura, T.~Miura, E.~Takasugi, and M.~Yoshimura, ``{Maximal CP violation,
  large mixings of neutrinos and democratic type neutrino mass matrix},''
  \href{http://dx.doi.org/10.1103/PhysRevD.61.073002}{{\em Phys. Rev.}
  {\bfseries D61} (2000) 073002},
\href{http://arxiv.org/abs/hep-ph/9909415}{{\ttfamily arXiv:hep-ph/9909415
  [hep-ph]}}.
%%CITATION = HEP-PH/9909415;%%.

\bibitem{Miura:2000sx}
T.~Miura, E.~Takasugi, and M.~Yoshimura, ``{Large CP violation, large mixings
  of neutrinos and the Z(3) symmetry},''
  \href{http://dx.doi.org/10.1103/PhysRevD.63.013001}{{\em Phys. Rev.}
  {\bfseries D63} (2001) 013001},
\href{http://arxiv.org/abs/hep-ph/0003139}{{\ttfamily arXiv:hep-ph/0003139
  [hep-ph]}}.
%%CITATION = HEP-PH/0003139;%%.

\bibitem{Ma:2002ce}
E.~Ma, ``{The All purpose neutrino mass matrix},''
  \href{http://dx.doi.org/10.1103/PhysRevD.66.117301}{{\em Phys. Rev.}
  {\bfseries D66} (2002) 117301},
\href{http://arxiv.org/abs/hep-ph/0207352}{{\ttfamily arXiv:hep-ph/0207352
  [hep-ph]}}.
%%CITATION = HEP-PH/0207352;%%.

\bibitem{Ma:2015fpa}
E.~Ma, ``{Neutrino mixing: $A_4$ variations},''
  \href{http://dx.doi.org/10.1016/j.physletb.2015.11.049}{{\em Phys. Lett.}
  {\bfseries B752} (2016) 198--200},
\href{http://arxiv.org/abs/1510.02501}{{\ttfamily arXiv:1510.02501 [hep-ph]}}.
%%CITATION = ARXIV:1510.02501;%%.

\bibitem{Ma:2016nkf}
E.~Ma, ``{Soft $A_4 \to Z_3$ symmetry breaking and cobimaximal neutrino
  mixing},'' \href{http://dx.doi.org/10.1016/j.physletb.2016.02.032}{{\em Phys.
  Lett.} {\bfseries B755} (2016) 348--350},
\href{http://arxiv.org/abs/1601.00138}{{\ttfamily arXiv:1601.00138 [hep-ph]}}.
%%CITATION = ARXIV:1601.00138;%%.

\bibitem{Damanik:2017jar}
A.~Damanik, ``{Neutrino masses from a cobimaximal neutrino mixing matrix},''
\href{http://arxiv.org/abs/1702.03214}{{\ttfamily arXiv:1702.03214
  [physics.gen-ph]}}.
%%CITATION = ARXIV:1702.03214;%%.

\bibitem{Ma:2017moj}
E.~Ma and G.~Rajasekaran, ``{Cobimaximal neutrino mixing from $A_4$ and its
  possible deviation},''
  \href{http://dx.doi.org/10.1209/0295-5075/119/31001}{{\em EPL} {\bfseries
  119} no.~3, (2017) 31001},
\href{http://arxiv.org/abs/1708.02208}{{\ttfamily arXiv:1708.02208 [hep-ph]}}.
%%CITATION = ARXIV:1708.02208;%%.

\bibitem{Ma:2017trv}
E.~Ma, ``{Cobimaximal neutrino mixing from $S_3 \times Z_2$},''
  \href{http://dx.doi.org/10.1016/j.physletb.2017.12.049}{{\em Phys. Lett.}
  {\bfseries B777} (2018) 332--334},
\href{http://arxiv.org/abs/1707.03352}{{\ttfamily arXiv:1707.03352 [hep-ph]}}.
%%CITATION = ARXIV:1707.03352;%%.

\bibitem{Grimus:2017itg}
W.~Grimus and L.~Lavoura, ``{Cobimaximal lepton mixing from soft symmetry
  breaking},'' \href{http://dx.doi.org/10.1016/j.physletb.2017.09.082}{{\em
  Phys. Lett.} {\bfseries B774} (2017) 325--331},
\href{http://arxiv.org/abs/1708.09809}{{\ttfamily arXiv:1708.09809 [hep-ph]}}.
%%CITATION = ARXIV:1708.09809;%%.

\bibitem{Ma:2019byo}
E.~Ma, ``{Two-loop $Z_4$ Dirac neutrino masses and mixing, with
  self-interacting dark matter},''
  \href{http://dx.doi.org/10.1016/j.nuclphysb.2019.114725}{{\em Nucl. Phys.}
  {\bfseries B946} (2019) 114725},
\href{http://arxiv.org/abs/1907.04665}{{\ttfamily arXiv:1907.04665 [hep-ph]}}.
%%CITATION = ARXIV:1907.04665;%%.

\bibitem{Das:2016czs}
D.~Das, M.~L. L\'opez-Ib\'a\~nez, M.~J. P\'erez, and O.~Vives, ``{Effective
  theories of flavor and the nonuniversal MSSM},''
  \href{http://dx.doi.org/10.1103/PhysRevD.95.035001}{{\em Phys. Rev. D}
  {\bfseries 95} no.~3, (2017) 035001},
  \href{http://arxiv.org/abs/1607.06827}{{\ttfamily arXiv:1607.06827
  [hep-ph]}}.

\bibitem{Lopez-Ibanez:2017xxw}
M.~L. L\'opez-Ib\'a\~nez, A.~Melis, M.~J. P\'erez, and O.~Vives, ``{Slepton
  non-universality in the flavor-effective MSSM},''
  \href{http://dx.doi.org/10.1007/JHEP11(2017)162}{{\em JHEP} {\bfseries 11}
  (2017) 162}, \href{http://arxiv.org/abs/1710.02593}{{\ttfamily
  arXiv:1710.02593 [hep-ph]}}. [Erratum: JHEP 04, 015 (2018)].

\bibitem{Lopez-Ibanez:2019rgb}
M.~L. L\'opez-Ib\'a\~nez, A.~Melis, D.~Meloni, and O.~Vives, ``{Lepton flavor
  violation and neutrino masses from A$_{5}$ and CP in the non-universal
  MSSM},'' \href{http://dx.doi.org/10.1007/JHEP06(2019)047}{{\em JHEP}
  {\bfseries 06} (2019) 047}, \href{http://arxiv.org/abs/1901.04526}{{\ttfamily
  arXiv:1901.04526 [hep-ph]}}.

\bibitem{Cabrera:2020lmg}
M.~E. Cabrera, J.~A. Casas, A.~Delgado, and S.~Robles, ``{2HDM singlet portal
  to dark matter},''
\href{http://arxiv.org/abs/2011.09101}{{\ttfamily arXiv:2011.09101 [hep-ph]}}.
%%CITATION = ARXIV:2011.09101;%%.

\bibitem{Varzielas:2016zjc}
I.~de~Medeiros~Varzielas, S.~F. King, C.~Luhn, and T.~Neder, ``{CP-odd
  invariants for multi-Higgs models: applications with discrete symmetry},''
  \href{http://dx.doi.org/10.1103/PhysRevD.94.056007}{{\em Phys. Rev.}
  {\bfseries D94} no.~5, (2016) 056007},
\href{http://arxiv.org/abs/1603.06942}{{\ttfamily arXiv:1603.06942 [hep-ph]}}.
%%CITATION = ARXIV:1603.06942;%%.

\bibitem{deMedeirosVarzielas:2017glw}
I.~de~Medeiros~Varzielas, S.~F. King, C.~Luhn, and T.~Neder, ``{Minima of
  multi-Higgs potentials with triplets of $\Delta(3n^2)$ and $\Delta(6n^2)$},''
  \href{http://dx.doi.org/10.1016/j.physletb.2017.11.005}{{\em Phys. Lett.}
  {\bfseries B775} (2017) 303--310},
\href{http://arxiv.org/abs/1704.06322}{{\ttfamily arXiv:1704.06322 [hep-ph]}}.
%%CITATION = ARXIV:1704.06322;%%.

\bibitem{Staub:2008uz}
F.~Staub, ``{SARAH},''
\href{http://arxiv.org/abs/0806.0538}{{\ttfamily arXiv:0806.0538 [hep-ph]}}.
%%CITATION = ARXIV:0806.0538;%%.

\bibitem{Staub:2013tta}
F.~Staub, ``{SARAH 4 : A tool for (not only SUSY) model builders},''
  \href{http://dx.doi.org/10.1016/j.cpc.2014.02.018}{{\em Comput. Phys.
  Commun.} {\bfseries 185} (2014) 1773--1790},
\href{http://arxiv.org/abs/1309.7223}{{\ttfamily arXiv:1309.7223 [hep-ph]}}.
%%CITATION = ARXIV:1309.7223;%%.

\bibitem{Porod:2014xia}
W.~Porod, F.~Staub, and A.~Vicente, ``{A Flavor Kit for BSM models},''
  \href{http://dx.doi.org/10.1140/epjc/s10052-014-2992-2}{{\em Eur. Phys. J.}
  {\bfseries C74} no.~8, (2014) 2992},
\href{http://arxiv.org/abs/1405.1434}{{\ttfamily arXiv:1405.1434 [hep-ph]}}.
%%CITATION = ARXIV:1405.1434;%%.

\bibitem{Goodsell:2015ira}
M.~Goodsell, K.~Nickel, and F.~Staub, ``{Generic two-loop Higgs mass
  calculation from a diagrammatic approach},''
  \href{http://dx.doi.org/10.1140/epjc/s10052-015-3494-6}{{\em Eur. Phys. J.}
  {\bfseries C75} no.~6, (2015) 290},
\href{http://arxiv.org/abs/1503.03098}{{\ttfamily arXiv:1503.03098 [hep-ph]}}.
%%CITATION = ARXIV:1503.03098;%%.

\bibitem{Staub:2015kfa}
F.~Staub, ``{Exploring new models in all detail with SARAH},''
  \href{http://dx.doi.org/10.1155/2015/840780}{{\em Adv. High Energy Phys.}
  {\bfseries 2015} (2015) 840780},
\href{http://arxiv.org/abs/1503.04200}{{\ttfamily arXiv:1503.04200 [hep-ph]}}.
%%CITATION = ARXIV:1503.04200;%%.

\bibitem{Goodsell:2018tti}
M.~D. Goodsell and F.~Staub, ``{Unitarity constraints on general scalar
  couplings with SARAH},''
  \href{http://dx.doi.org/10.1140/epjc/s10052-018-6127-z}{{\em Eur. Phys. J.}
  {\bfseries C78} no.~8, (2018) 649},
\href{http://arxiv.org/abs/1805.07306}{{\ttfamily arXiv:1805.07306 [hep-ph]}}.
%%CITATION = ARXIV:1805.07306;%%.

\bibitem{Porod:2003um}
W.~Porod, ``{SPheno, a program for calculating supersymmetric spectra, SUSY
  particle decays and SUSY particle production at e+ e- colliders},''
  \href{http://dx.doi.org/10.1016/S0010-4655(03)00222-4}{{\em Comput. Phys.
  Commun.} {\bfseries 153} (2003) 275--315},
\href{http://arxiv.org/abs/hep-ph/0301101}{{\ttfamily arXiv:hep-ph/0301101
  [hep-ph]}}.
%%CITATION = HEP-PH/0301101;%%.

\bibitem{Porod:2011nf}
W.~Porod and F.~Staub, ``{SPheno 3.1: Extensions including flavour, CP-phases
  and models beyond the MSSM},''
  \href{http://dx.doi.org/10.1016/j.cpc.2012.05.021}{{\em Comput. Phys.
  Commun.} {\bfseries 183} (2012) 2458--2469},
\href{http://arxiv.org/abs/1104.1573}{{\ttfamily arXiv:1104.1573 [hep-ph]}}.
%%CITATION = ARXIV:1104.1573;%%.

\bibitem{Xing:2019vks}
Z.-z. Xing, ``{Flavor structures of charged fermions and massive neutrinos},''
  \href{http://dx.doi.org/10.1016/j.physrep.2020.02.001}{{\em Phys. Rept.}
  {\bfseries 854} (2020) 1--147},
\href{http://arxiv.org/abs/1909.09610}{{\ttfamily arXiv:1909.09610 [hep-ph]}}.
%%CITATION = ARXIV:1909.09610;%%.

\bibitem{Zyla:2020zbs}
{\bfseries Particle Data Group} Collaboration, P.~A. Zyla {\em et~al.},
  ``{Review of Particle Physics},''
\href{http://dx.doi.org/10.1093/ptep/ptaa104}{{\em PTEP} {\bfseries 2020}
  no.~8, (2020) 083C01}.
%%CITATION = INSPIRE-1812251;%%.

\bibitem{deSalas:2020pgw}
P.~F. de~Salas, D.~V. Forero, S.~Gariazzo, P.~Martínez-Miravé, O.~Mena, C.~A.
  Ternes, M.~Tórtola, and J.~W.~F. Valle, ``{2020 Global reassessment of the
  neutrino oscillation picture},''
\href{http://arxiv.org/abs/2006.11237}{{\ttfamily arXiv:2006.11237 [hep-ph]}}.
%%CITATION = ARXIV:2006.11237;%%.

\bibitem{Esteban:2020cvm}
I.~Esteban, M.~C. Gonzalez-Garcia, M.~Maltoni, T.~Schwetz, and A.~Zhou, ``{The
  fate of hints: updated global analysis of three-flavor neutrino
  oscillations},'' \href{http://dx.doi.org/10.1007/JHEP09(2020)178}{{\em JHEP}
  {\bfseries 09} (2020) 178},
\href{http://arxiv.org/abs/2007.14792}{{\ttfamily arXiv:2007.14792 [hep-ph]}}.
%%CITATION = ARXIV:2007.14792;%%.

\bibitem{Armengaud:2020luj}
{\bfseries CUPID} Collaboration, E.~Armengaud {\em et~al.}, ``{A new limit for
  neutrinoless double-beta decay of $^{100}$Mo from the CUPID-Mo experiment},''
  \href{http://arxiv.org/abs/2011.13243}{{\ttfamily arXiv:2011.13243
  [nucl-ex]}}.

\bibitem{Albert:2017hjq}
{\bfseries nEXO} Collaboration, J.~B. Albert {\em et~al.}, ``{Sensitivity and
  Discovery Potential of nEXO to Neutrinoless Double Beta Decay},''
  \href{http://dx.doi.org/10.1103/PhysRevC.97.065503}{{\em Phys. Rev. C}
  {\bfseries 97} no.~6, (2018) 065503},
  \href{http://arxiv.org/abs/1710.05075}{{\ttfamily arXiv:1710.05075
  [nucl-ex]}}.

\bibitem{Barabash:2019suz}
A.~S. Barabash, ``{Possibilities of future double beta decay experiments to
  investigate inverted and normal ordering region of neutrino mass},''
  \href{http://dx.doi.org/10.3389/fphy.2018.00160}{{\em Front. in Phys.}
  {\bfseries 6} (2019) 160}, \href{http://arxiv.org/abs/1901.11342}{{\ttfamily
  arXiv:1901.11342 [nucl-ex]}}.

\bibitem{Bertl:2006up}
{\bfseries SINDRUM II} Collaboration, W.~H. Bertl {\em et~al.}, ``{A Search for
  muon to electron conversion in muonic gold},''
  \href{http://dx.doi.org/10.1140/epjc/s2006-02582-x}{{\em Eur. Phys. J. C}
  {\bfseries 47} (2006) 337--346}.

\bibitem{Bernstein:2019fyh}
{\bfseries Mu2e} Collaboration, R.~Bernstein, ``{The Mu2e Experiment},''
  \href{http://dx.doi.org/10.3389/fphy.2019.00001}{{\em Front. in Phys.}
  {\bfseries 7} (2019) 1}, \href{http://arxiv.org/abs/1901.11099}{{\ttfamily
  arXiv:1901.11099 [physics.ins-det]}}.

\bibitem{Kuno:2013mha}
{\bfseries COMET} Collaboration, Y.~Kuno, ``{A search for muon-to-electron
  conversion at J-PARC: The COMET experiment},''
  \href{http://dx.doi.org/10.1093/ptep/pts089}{{\em PTEP} {\bfseries 2013}
  (2013) 022C01}.

\bibitem{Barlow:2011zza}
R.~Barlow, ``{The PRISM/PRIME project},''
  \href{http://dx.doi.org/10.1016/j.nuclphysbps.2011.06.009}{{\em Nucl. Phys. B
  Proc. Suppl.} {\bfseries 218} (2011) 44--49}.

\bibitem{Deppisch:2015qwa}
F.~F. Deppisch, P.~S. Bhupal~Dev, and A.~Pilaftsis, ``{Neutrinos and Collider
  Physics},'' \href{http://dx.doi.org/10.1088/1367-2630/17/7/075019}{{\em New
  J. Phys.} {\bfseries 17} no.~7, (2015) 075019},
  \href{http://arxiv.org/abs/1502.06541}{{\ttfamily arXiv:1502.06541
  [hep-ph]}}.

\bibitem{Ishimori:2010au}
H.~Ishimori, T.~Kobayashi, H.~Ohki, Y.~Shimizu, H.~Okada, and M.~Tanimoto,
  ``{Non-Abelian Discrete Symmetries in Particle Physics},''
  \href{http://dx.doi.org/10.1143/PTPS.183.1}{{\em Prog. Theor. Phys. Suppl.}
  {\bfseries 183} (2010) 1--163},
\href{http://arxiv.org/abs/1003.3552}{{\ttfamily arXiv:1003.3552 [hep-th]}}.
%%CITATION = ARXIV:1003.3552;%%.

\end{thebibliography}
\end{document}